\documentclass[11pt]{article}
\usepackage{graphicx} 

 \pdfoutput=1

 \usepackage{axodraw4j}
\usepackage{pstricks}
\usepackage{mathtools}

 \usepackage{tikz}
\usetikzlibrary{decorations.pathmorphing}
\usetikzlibrary{patterns}
\usetikzlibrary{cd}
\usetikzlibrary{snakes}
\usepackage{latexsym,amsmath,amsfonts,amssymb,amsthm}
\usepackage{mathrsfs}
\usepackage{enumitem}
\usepackage[latin1]{inputenc}
\usepackage[american]{babel}
\usepackage{bbm}
\usepackage{cite}
\usepackage{tikz-feynman}
\usepackage{amsfonts}
\usepackage{graphicx}
\usepackage{amsmath,bm}
\usepackage{amssymb}
\usepackage{latexsym}
\usepackage{mathrsfs}
\usepackage{slashed}
\definecolor{red}{rgb}{1,0,0}
\usepackage{cancel}

\usepackage{soul} 

\usepackage{setspace} 

\usepackage{subcaption}

\definecolor{red}{rgb}{1,0,0}

\makeatletter
\def\section{\@startsection {section}{1}{\z@}{-3.5ex plus -1ex minus
 -.2ex}{2.3ex plus .2ex}{\large\bf}}
\def\subsection{\@startsection{subsection}{2}{\z@}{-3.25ex plus -1ex
minus -.2ex}{1.5ex plus .2ex}{\normalsize\bf}}
\makeatother
\makeatletter

\@addtoreset{equation}{section}

\makeatother

\newcommand{\CO}{{\cal O}}

\def\bel{\begin{equation}\begin{aligned}}
\def\eel{\end{aligned}\end{equation}}

\def\bea{\begin{eqnarray}} \def\eea{\end{eqnarray}}
\def\be{\begin{equation}} \def\ee{\end{equation}} 
\def\nn{\nonumber}

\newcommand{\promille}{%
  \relax\ifmmode\promillezeichen
        \else\leavevmode\(\mathsurround=0pt\promillezeichen\)\fi}
\newcommand{\promillezeichen}{%
  \kern-.05em%
  \raise.5ex\hbox{\the\scriptfont0 0}%
  \kern-.15em/\kern-.15em%
  \lower.25ex\hbox{\the\scriptfont0 00}}

\newcommand{\re}{{\rm e}}

\newcommand{\OO}{\mathcal{O}}

\usepackage[colorlinks,linkcolor=black,citecolor=blue,urlcolor=blue,linktocpage]{hyperref}

\begin{document}
\setcounter{page}{0}
\thispagestyle{empty}

\parskip 3pt

\font\mini=cmr10 at 2pt

\begin{titlepage}
~\vspace{1cm}
\begin{center}

	\vspace*{-.6cm}

	\begin{center}

		\vspace*{1.1cm}

		{\centering \Large\textbf{An anharmonic alliance: exact WKB meets EPT}}

	\end{center}

	\vspace{0.8cm}
	{\bf Bruno Bucciotti$^{a,d}$, Tomas Reis$^{b,c}$, and Marco Serone$^{b,c}$}

	\vspace{1.cm}
	
	${}^a\!\!$
	{\em  Scuola Normale Superiore, Piazza dei cavalieri 7 Pisa, Italy}
		
	\vspace{.3cm}

	${}^b\!\!$
		{\em SISSA, Via Bonomea 265, I-34136 Trieste, Italy}
			
			\vspace{.3cm}
	${}^c\!\!$			
	{\em INFN, Sezione di Trieste, Via Valerio 2, I-34127 Trieste, Italy}
 
			\vspace{.3cm}
	${}^d\!\!$			
	{\em INFN, Sezione di Pisa, Largo B. Pontecorvo, 3, I-56127 Pisa, Italy}

\vskip 5pt

{\em \small{bruno.bucciotti@sns.it,serone@sissa.it,treis@sissa.it}}

\end{center}

\begin{abstract}

Certain quantum mechanical systems with a discrete spectrum, whose observables are given by a transseries in $\hbar$, were shown to admit
$\hbar_0$-deformations with Borel resummable expansions which reproduce the original model at $\hbar_0=\hbar$.
Such expansions were dubbed Exact Perturbation Theory (EPT).
We investigate how the above results can be obtained within the framework of the exact WKB 
method by studying the spectrum of polynomial
quantum mechanical systems.
Within exact WKB, energy eigenvalues are determined by exact quantization conditions defined in terms of 
Voros symbols $a_{\gamma_i}$, $\gamma_i$ being their associated cycles, and generally give rise to transseries in $\hbar$. After reviewing how the Borel summability of energy eigenvalues in the quartic anharmonic potential emerges in exact WKB, we extend it to higher order anharmonic potentials with quantum corrections. 
We then show that any polynomial potential can be $\hbar_0$-deformed to a model where the exact quantization condition reads simply $a_\gamma=-1$ 
and leads to the EPT  Borel resummable series for all energy eigenvalues.

\end{abstract}

\end{titlepage}

\tableofcontents

\section{Introduction}

Quantum mechanical models have been, and still are, an ideal playground to understand the nature of perturbation theory.
Thanks to the Schr\"odinger equation we have easy access to perturbative terms at large orders. Starting from 
the seminal papers by Bender and Wu \cite{Bender:1969si,Bender:1973rz}, this allowed us to get accurate estimates of large order behaviours of perturbative asymptotic series.
Famously, the Schr\"odinger equation can also be studied using a Wentzel--Kramers--Brillouin (WKB) approximation \cite{Wentzel:1926aor,Kramers:1926njj,Brillouin:1926blg} which in fact played a key role also in \cite{Bender:1969si,Bender:1973rz}. The proper use of the WKB approximation was found much later \cite{voros-quartic,silverstone} using 
Borel resummation and resurgence techniques \cite{ecalle}. In particular, building on previous works \cite{Balian:1974ah,bpv1}, Voros upgraded the WKB approximation
into an exact method, exact WKB (EWKB), which was further developed and laid on solid mathematical foundation in \cite{AKT1,reshyper,ddpham,dpham,AKT2}.
In EWKB, energy eigenvalues of a quantum mechanical system are determined by exact quantization conditions (EQCs) defined in terms of so called
Voros symbols $a_{\gamma_i}$, where $\gamma_i$ are the associated periods in complexified space between the turning points of the classical potential.  
In certain quantum mechanical models EQCs were guessed based on truncated transseries in multi-instantons computations \cite{Zinn-Justin:1982hva}, and then rigorously established using EWKB \cite{ddpham}. In general, EQCs involve both classically allowed (perturbative) and disallowed (non-perturbative) periods
and generally give rise to transseries in $\hbar$, $\exp(-1/\hbar)$ and possibly $\log \hbar$. 

Several connections between EWKB and other subjects have been worked out. These include connections with ${\cal N}=2$ supersymmetric gauge theories \cite{Gaiotto:2009hg,Mironov:2009uv}, topological strings \cite{cm-ha,coms}, thermodynamic Bethe ansatz  \cite{Ito:2018eon}
via the ordinary differential equations/integrable models (ODE/IM) correspondence \cite{dt}, generalizations in terms of the geometry of quantum periods \cite{Gu:2022fss} and so on.
It is fair to say that EWKB represents one of the most interesting and established applications of resurgence in theoretical physics.
How to unpack the EQCs to efficiently write transseries for energy eigenvalues or other observables, and their relation to transseries coming from multi-intantons in a path integral approach, has also been the subject of some activity in the last years, see e.g.  \cite{dunne-unsal,Basar:2017hpr,Sueishi:2020rug,Sueishi:2021xti,Kamata:2021jrs}.

Independently of EWKB it has been shown that several one-dimensional quantum mechanical models with a discrete spectrum admit (in general infinite) $\hbar_0$-deformations (the original system is recovered by setting $\hbar_0=\hbar$) such that the path integral is exactly determined by a single saddle-point \cite{Serone:2016qog,power}.
If we Borel resum the $\hbar_0$-deformed perturbative series in $\hbar$ at {\it fixed} $\hbar_0$
and {\it after} we set $\hbar_0=\hbar$, the exact result is recovered. The ``exact perturbation theory'' (EPT) in the $\hbar_0$-model is then able to capture the full result in models which are known to receive instanton corrections, such as the (supersymmetric) double well. This somewhat surprising and powerful result was obtained using path integrals and steepest-descent methods (Lefschetz thimbles). 

The aim of this paper is to understand how EPT emerges from an EWKB analysis. 

We start in section \ref{sec:basics} with a brief review on EWKB, with an emphasis on how to determine EQCs. 
In general, EQCs are a constraint of the form
\be
F[a_{\gamma_i}(E) ] = 0\,,
\label{eq:Fexsol}
\ee
where 
\begin{equation}
a_{\gamma_i} = \exp\Big( \frac{1}{\hbar}\oint_{\gamma_i} P_{\text{even}} (z) dz \Big)
\label{eq:VorosDef}
\end{equation}
are the Voros symbols associated to the different cycles connecting classical turning points, and $P_{\text{even}}$ is the Borel resummation of the $\hbar$ series starting with 
the classical contribution $\sqrt{2(V-E)}$ followed by quantum corrections. At fixed parameters of the potential $V$, the Voros symbols depend only on the energy $E$
entering the Schr\"odinger equation \eqref{eq:SchroC}. Energy eigenvalues can be determined exactly as those values for which \eqref{eq:Fexsol} is satisfied.
However, if one is interested in determining their associated asymptotic series (or transseries) in $\hbar$, \eqref{eq:Fexsol} should be ``downgraded" to its formal power series form $\widetilde F$. In $\widetilde F$ we replace $E$ by a transseries $\widetilde E$ which is determined by demanding  $\widetilde F=0$
order by order in $\hbar$, and possibly in the transseries parameters $\exp(-1/\hbar)$ and $\log \hbar$. 
Importantly, EQCs are not uniquely determined and depend on $E$ and $\arg\, \hbar$, because Voros symbols are subject to Stokes jumps.

In section \ref{sec:pure-anharmonic} we study in some detail the EQCs in the quartic anharmonic oscillator as a function of $\arg\hbar$ and uncover a Stokes jump occurring for small $\arg \hbar$ when
two turning points approach each other in the limit $E\rightarrow 0$. 
Because of this discontinuity, the limits $\arg \hbar\rightarrow 0$ and $E\rightarrow 0$ do not commute. 
By taking $E\rightarrow 0$ first, we get
the simple EQC
\be
a_\gamma = -1\,,
\label{eq:EQCsimple}
\ee
where $\gamma$ is the perturbative cycle, which is shown to be Borel resummable and to lead to Borel resummable energy eigenvalues for sufficiently small $\hbar$, reproducing in this way a result of \cite{ddpham} but bypassing the complications related to double turning points.  
This analysis is then generalized to more general anharmonic potentials of the form \eqref{eq:PotAnhGen} in section
\ref{subsec:anharm-higher}.

Using EPT, which we briefly review in \ref{subsec:ept}, any bounded polynomial potential can be reduced to the form \eqref{eq:PotAnhGen} plus a quantum potential which includes the remaining terms. We then prove in sections \ref{sec:quant-anharm} and \ref{subsec:LOBq} that energy eigenvalues of an arbitrary quantum mechanical model with a bounded polynomial potential admit $\hbar_0$-deformed EPT series which are Borel resummable with EQC given by \eqref{eq:EQCsimple}. We conclude in section \ref{sec:conclu}.
We report in appendices \ref{app:connmat} and \ref{app:defquadratic} the derivation of so called connection matrices for simple and double turning points, which are important
ingredients to determine EQCs. In appendix \ref{app:purequartic} we discuss the transseries associated to the energy eigenvalues in the pure quartic model (with no use of EPT) and
prove that the radius of convergence of the partial series of exponential corrections is finite.

A note on notation. We carefully distinguish actual functions from their formal asymptotic power series expression by putting a tilde on the latter. 
A hat denotes the associated Borel function:  
\be
\widetilde f = \sum_{n=0} c_n \hbar^n \,, \qquad  \hat f = \sum_{n=0} \frac{c_n}{n!} t^n \,.
\ee
If $\widetilde f$ is a Gevrey-1 series (i.e. $c_n\sim (n!)^1$ for large $n$), then $\hat f$ is analytic in the Borel $t$-plane in a disc around the origin and is analytically continuable in the Borel plane. We define the Laplace transform in the direction $\theta$ as
\be
f^\theta(\hbar) = s_\theta(\widetilde f) =\frac{1}{\hbar}  \int_0^{e^{i \theta} \infty}\!\! \!\!dt \, \hat f(t) e^{-\frac{t}{\hbar}}\,.
\label{eq:BorelDef}
\ee
If $|\hat f(t e^{i \theta})|\leq e^{a t}$ for any $t\geq 0$, then $f^\theta(\hbar)$ is analytic for Re$(e^{i \theta}\hbar^{-1}) > a$.
If the ray $\theta=0$ is not a Stokes line, then $f^0$ is well-defined and might reconstruct the original function $f$.
In the complex $\hbar$-plane we often have wedges, delimited by Stokes lines, labelled by roman numbers $\text{I, II},\ldots$.
We then write
\be
f^\text{N} = f^\theta\,, \qquad \theta \in \text{wedge} \; \text{N}, \qquad \text{N}=\text{I,II},\ldots \,.
\ee

\section{Exact WKB basics}
\label{sec:basics}

In this section we briefly review basics of EWKB.
As mentioned in the introduction, the spectrum of a given quantum mechanical system is encoded in the Voros symbols \eqref{eq:VorosDef} and 
energy eigenvalues are given by the solutions of the EQC of the form \eqref{eq:Fexsol}.  
Wave functions do not enter in EQCs, but the determination of the latter requires an understanding of the former.
We then review the EWKB rules to derive EQCs. We discuss the ansatz for the wave function $\psi(z)$ in section \ref{subsec:ansatz}, 
review how the $z$-plane splits into so called Stokes region and how Voros symbols and wave functions jump in section \ref{subsec:regions}, and finally the rules of how to determine EQCs for an arbitrary polynomial potential 
in the case in which all its zeros 
are simple in section \ref{subsec:EQCs}. We report in appendix \ref{app:connmat}
the computation of connection matrices for pure monomial potentials.
Determining EQCs in presence of higher order zeros is more complicated and there are no general rules as for simple turning points. 
Nevertheless we report in appendix \ref{app:defquadratic} the connection matrices for a deformed quadratic turning point,
which will be useful in our considerations. There are no new results in this section, which contains standard material, though some emphasis might differ with respect to other presentations. More details on EWKB can be found e.g. in \cite{INcluster}. For a more elementary textbook presentation see \cite{marino_2021}.

\subsection{Ansatz for the wave function}
\label{subsec:ansatz}

The starting point of a WKB analysis is the (complexified) Schr\"odinger equation
\be
-\hbar^2 \frac{d^2\psi(z)}{dz^2} + Q(z) \psi(z) = 0 \,,
\label{eq:SchroC}
\ee
where $\hbar, z\in \mathbf{C}$ and 
\be
Q(z) = 2(V(z)-E)\,. 
\label{eq:Qdef}
\ee
For simplicity we assume in the following a potential $V$ given by an entire function, where the only singularity is at $z=\infty$.
In general the potential $V$ could be a quantum potential, namely it can depend on $\hbar$:
\be
Q(z,\hbar) = \sum_{n=0}^\infty Q_n(z) \hbar^n\,.
\ee
The fundamental WKB ansatz is
\be
\psi(z) = c e^{\frac{1}{\hbar} \int^z_{z_0} P(w,\hbar) dw}\,,
\label{eq:ansatz}
\ee
where $c$ is a constant, which could be reabsorbed in the definition of $z_0$, but we prefer to keep it explicitly.
Plugging \eqref{eq:ansatz} in \eqref{eq:SchroC} gives rise to a Riccati equation for $P$:
\be
(\hbar P' + P^2 - Q ) = 0\,,
\label{eq:Riccati}
\ee
where $f'\equiv df/dz$. We look for asymptotic solutions of $P$ in the form
\be
\widetilde P=\sum_{n=0}^\infty P_n \hbar^n\,.
\label{eq:Pexp}
\ee
The coefficient functions $P_n(z)$ satisfy the following recursion relations:
\be
2 P_0 P_{n+1} =  Q_{n+1}- P_n' - \sum_{k=1}^n P_k P_{n+1-k} \,, \qquad n\geq 0\,.
\label{eq:recrel}
\ee
Once $P_0$ is found, the whole series is fixed by \eqref{eq:recrel}.
We then get two series, depending on which classical term $P_0$ is selected:  
\be
P_0^\eta = \pm \sqrt{Q_0}\,,  \qquad   \eta = \pm 1 \,. 
\ee
The first two terms beyond $P_0^\eta$ read
\begin{align}
P_1^\eta & = - \frac{Q_0'}{4Q_0} + \eta  \frac{Q_1}{2 \sqrt{Q_0}} \,, \label{eq:Pcoeff} \\
P_2^\eta & = \frac{\eta}{32Q_0^{\frac 52}}  \Big(-5 Q_0^{'2} +4 Q_0 Q_0'' - 4 Q_0 Q_1^2 + 16 Q_0^2 Q_2 \Big)
+ \frac {Q_1 Q_0'-Q_0 Q_1'}{4Q_0^2} \,. \nn
\end{align}
When $Q=Q_0$ (classical potential only) we have
\be
P_{2n+1}^- =  P_{2n+1}^+ \,, \qquad P_{2n}^- = - P_{2n}^+\,, \qquad n\geq 0\,.
\label{ew:PpmRel}
\ee
For now let us assume $Q=Q_0$ and write
\be
\begin{split}
\widetilde P_{\text{even}} & =  \frac{\widetilde P^+-\widetilde P^-}{2} = \sum_{n=0}^\infty \hbar^{2n} P_{2n} \,,\qquad \qquad \qquad (Q=Q_0) \\ 
\widetilde P_{\text{odd}} &=  \frac{\widetilde P^++\widetilde  P^-}{2}  = \sum_{n=0}^\infty \hbar^{2n+1} P_{2n+1}\,, \qquad \qquad (Q=Q_0) 
\label{eq:Pevenodd}
\end{split}
\ee
We plug \eqref{eq:Pevenodd} in \eqref{eq:Riccati} and split even and odd terms in $\hbar$:
\begin{align}
\hbar \widetilde P'_{\text{even}} + 2 \widetilde P_{\text{even}} \widetilde P_{\text{odd}}  & =0 \,, \\
\hbar \widetilde P'_{\text{odd}} +  \widetilde P_{\text{even}}^2+ \widetilde P_{\text{odd}}^2 - Q_0 & =0 \,.
\end{align}
The first equation gives
\be
\widetilde P_{\text{odd}} = -\frac{\hbar}{2} \frac{d \log \widetilde P_{\text{even}}}{dz} \,,
\label{eq:PoddasPeven}
\ee
and hence
\be
\widetilde \psi_\pm = c \sqrt{\frac{\widetilde P_{\text{even}}(z_0)}{\widetilde P_{\text{even}}(z)}} e^{\pm \frac{1}{\hbar} \int^z_{z_0} \widetilde P_{\text{even}} dw}   \,.
\label{eq:psiasPevenAux}
\ee
For generic $Q$, \eqref{ew:PpmRel} does not apply, but we can still define $\widetilde P_{\text{even}}$ and $\widetilde P_{\text{odd}}$ as  
\be
\begin{split}
\widetilde P_{\text{even}} & =  \frac{\widetilde P^+-\widetilde P^-}{2} = \sum_{n=0}^\infty \hbar^{n} P_{\text{even},n} \,, \\ 
\widetilde P_{\text{odd}} &=  \frac{\widetilde P^++\widetilde  P^-}{2}  = \sum_{n=1}^\infty \hbar^{n} P_{\text{odd},n}\,.
\label{eq:PevenoddV2}
\end{split}
\ee
They can no longer be expressed as even and odd powers of $\hbar$, respectively, but they still satisfy \eqref{eq:PoddasPeven}.
Indeed, the Riccati equations $(\hbar \widetilde P^{'\pm} + (\widetilde P^{\pm})^2 - Q ) = 0$ turn into
\begin{align}
\hbar \widetilde P'_{\text{even}}+\hbar \widetilde P'_{\text{odd}}  + (\widetilde P_{\text{even}}+\widetilde P_{\text{odd}})^2 - Q  &= 0  \,, \nn \\
-\hbar \widetilde P'_{\text{even}}+\hbar \widetilde P'_{\text{odd}}  + (\widetilde P_{\text{even}}-\widetilde P_{\text{odd}})^2 - Q & = 0 \,,
\end{align}
and the difference between the two equations gives \eqref{eq:PoddasPeven}. We then learn that \eqref{eq:psiasPevenAux} applies also for {\it general} quantum potentials $Q$.
We will generally choose $z_0$ to be a (simple) turning point and 
\be
c = \frac{1}{\sqrt{\widetilde P_{\text{even}}(z_0)}}\,,
\ee
so that 
\be
\widetilde \psi_\pm(z) = \frac{1}{\sqrt{\widetilde P_{\text{even}}(z)}} e^{\pm \frac{1}{\hbar} \int^z_{z_0} \widetilde P_{\text{even}}(w) dw}   \,.
\label{eq:psiasPeven}
\ee

\subsection{Stokes lines and periods}
\label{subsec:regions}

The solutions \eqref{eq:psiasPeven} of the Schr\"odinger equation are only formal asymptotic series which require resummation. 
In general we cannot have a unique resummation which applies globally. This can be seen by noting that while the actual solution has to be single valued in the $z$-plane for sufficiently regular potentials with a lower bound, in general \eqref{eq:psiasPeven} are not. Stokes phenomena occur and we have to determine the relations between the Borel resummations of \eqref{eq:psiasPeven} in different sectors of the complex plane. 
As we will see, such relations are encoded in $2\times 2$  connection matrices.
Stokes phenomena are determined by the zeroes and poles of $Q_0$. 
We mostly focus on polynomial potentials where the only pole is at infinity, while we can have simple or higher-order zeros named turning points.
Let us consider the generic situation of a point in moduli space (parameters of the potential and the energy $E$) where $Q_0$ has only simple turning points. 
The complex $z$-plane is divided in sectors delimited by Stokes lines defined as
\begin{equation}
\text{Im} \left\{\frac{1}{\hbar}\int_{\text{A}}^z \sqrt{Q_0(w) } dw\right\} = 0\,,
\label{eq:StokesLineDef}
\end{equation}
where A denotes a simple turning point of $Q_0$. Stokes lines are denoted regular if they start at point A and end at infinity, and singular if they start at A
and end at another turning point B.\footnote{When $Q_0$ has poles (in addition to the one at infinity) more Stokes trajectories are possible. See e.g. \cite{INcluster} for a clear exposition.}
Configurations with singular Stokes lines lead to ambiguities and should be avoided by a proper deformation. 
We will generally avoid singular Stokes lines by assigning a phase to $\hbar$. 
 
The key objects in EWKB are the so called periods, which are integrals between two turning points A and B. 
The various turning points of the potential makes generally the plane $z$ into a Riemann surface ${\cal M}$ of genus $g$ (depending on the potential). The contours around turning points are non-trivial cycles in $H^1({\cal M},\mathbb{C})$. The classical period is defined as
\begin{equation}
\Pi_{0,AB} = 2\int_A^B  \sqrt{Q_0(z) } dz =  \oint_\gamma  \sqrt{Q_0(z) } dz \,,
\label{eq:Pi0}
\end{equation}
where $\gamma$ is a cycle encircling the points A and B.\footnote{As it is evident from \eqref{eq:Pi0}, a square root branch-cut emanates from a simple turning point, so the contour integral does not vanish.} 
The quantum periods are defined as
\begin{equation}
\Pi_{AB} = s \Big(\widetilde \Pi_{AB}\Big)\,, \qquad \widetilde \Pi_{AB}= 2\int_A^B \widetilde P_{\text{even}}(z) dz =  \oint_\gamma \widetilde P_{\text{even}} (z) dz \,,
\label{eq:qPeriod}
\end{equation}
where $s$ denotes Borel resummation in the appropriate wedge of the complex plane where $\gamma$ sits. 
For each period we define the Voros symbol
\begin{equation}
a_{AB} = \re^{ \frac{1}{\hbar} \Pi_{AB}} = a_{BA}^{-1}\,.
\label{eq:VorosDef2}
\end{equation}
Sometimes we use the notation $\Pi_\gamma$ and $a_\gamma$ instead of $\Pi_{AB}$ and $a_{AB}$. Note that the definition of periods is done unambiguously in the Riemann surface, but the labelling of periods through the turning points pressuposes a choice of principal sheet to specify the orientation. 

If the Stokes lines are all regular, all the quantum periods $\Pi_{AB}$ are Borel resummable and well-defined.\footnote{If a given period happens to cross a regular Stokes line, we can decompose it in terms of products of periods 
defined in wedges without Stokes lines intersection.} As we vary the phase of $\hbar$, or move in moduli space, singular Stokes lines can emerge and lead to a change of configurations of Stokes lines in the $z$-plane. Singular Stokes lines can be seen as periods themselves as they start and end at simple turning points. Let $\gamma$ be the cycle associated to a singular Stokes line and $\alpha$ be the phase of the associated classical period:
\begin{equation}
\alpha\equiv \arg \Pi_{\gamma,0} =\arg\oint_\gamma \sqrt{Q_0(z) }dz  \,.
\label{arg-Pi}
\end{equation}
According to \eqref{eq:StokesLineDef}, the period \eqref{arg-Pi} can correspond to a (singular) Stokes line only for $\hbar  = |\hbar| {\re^{i \alpha}}$.
We can deform from the singular configuration by changing the phase of $\hbar$ as $\alpha\rightarrow \alpha\pm = \alpha\pm \epsilon$, with $0<\epsilon\ll 1$. The Voros symbols $a_\lambda$ of a cycle $\lambda$ computed in the configurations $\alpha-$ and $\alpha+$ are not the same, but related through the celebrated Dillinger-Delabaere-Pham (DDP) formula \cite{reshyper,dpham}
\begin{equation}
s_{\alpha-} (\widetilde a_\lambda) = \prod_{\arg \Pi_{\gamma_i,0}=\alpha} \left(1+ s_{\alpha+}(\widetilde a_{\gamma_i}^{-1})\right)^{-\langle\gamma_i,\lambda\rangle}s_{\alpha+}(\widetilde a_\lambda).
\label{eq:DDP}
\end{equation} 
In \eqref{eq:DDP} the product runs over all possible singular cycles with associated classical period of phase $\alpha$ and $\langle\gamma_i,\lambda\rangle$ denotes the intersection number of the two cycles.\footnote{The intersection number is  topological and thus rotational invariant, anti-symmetric between the two cycles and swaps sign if the orientation of a cycle is reversed. 
We take
\begin{equation}
\langle \rightarrow,\uparrow\rangle = +1,\quad 
\langle \leftarrow,\uparrow\rangle = -1,\quad 
\langle \rightarrow,\downarrow\rangle = -1,\quad 
\langle \leftarrow,\downarrow\rangle = +1. 
\end{equation}}
If $\gamma$ is a singular Stokes line at $\alpha$, the opposite cycle is a singular Stokes line at $\pi+\alpha$:
\be 
\arg \Pi_{\gamma^{-1},0} =  \arg \Pi_{\gamma,0} + \pi\,.
\label{eq:StokesInv}
\ee
The jump of $a_\lambda$ is different, because in the r.h.s. of \eqref{eq:DDP} $\widetilde a_{\gamma_i}^{-1} \rightarrow \widetilde a_{\gamma_i}$
and $\langle\gamma_i,\lambda\rangle\rightarrow - \langle\gamma_i,\lambda\rangle$.
Without loss of generality we can then restrict the range of $\alpha$ in \eqref{eq:DDP} to $-\pi/2< \alpha \leq \pi/2$. 
Note that the $\widetilde a_{\gamma_i}$ appear as purely non-perturbative corrections since $\widetilde a_{\gamma}^{-1}\sim \re^{-\frac{1}{\hbar}\Pi_{\gamma,0}}(1+\cdots)$ and $\frac{1}{\hbar}\Pi_{\gamma,0}$ is positive and real. Moreover, unless a singular Stokes cycle $\gamma$ is non-trivially linked with another singular cycle, \eqref{eq:DDP} implies that $s_{\alpha+} (\widetilde a_\gamma) = s_{\alpha-} (\widetilde a_\gamma)$, since  $\langle\gamma,\gamma\rangle=0$.
We will make extensive use of \eqref{eq:DDP} in this paper.

\begin{figure}[t!]
    \centering
    \raisebox{-3.em}
 {\scalebox{1.4}{
    \begin{tikzpicture}
\draw[snake=snake,segment length=5pt] (-1,3)-- (1,3); 

   \filldraw[fill=black, draw=black]  (1,3) circle (0.05 cm);

       \draw [stealth-] (2,3) - - (3,3);
            \draw (1,3) - - (2,3);
              \draw [-stealth]  (1.,3.) to (0.5,3.866);
         \draw (0.5,3.866) to (0,4.732);
      \draw [-stealth]  (1.,3.) to (0.5,2.134);
           \draw (0.5,2.134) to (0,1.268);
 \end{tikzpicture}}}
    \caption{Stokes lines and their orientation for the Airy function. The wavy black line represents the branch-cut.} 
	\label{fig:Airy}

\end{figure}
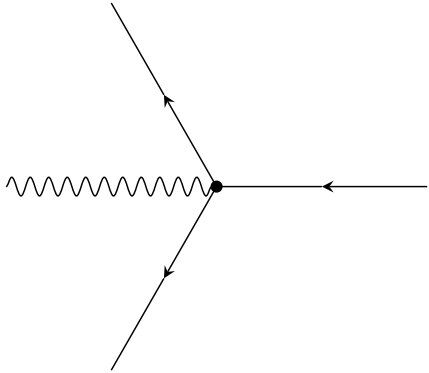

\begin{figure}[t!]
    \centering
    \raisebox{-3.em}
 {\scalebox{1.4}{
    \begin{tikzpicture}

   \filldraw[fill=black, draw=black]  (1,3) circle (0.05 cm);
      \draw [stealth-] (2,3) - - (3,3);
            \draw (1,3) - - (2,3);
          \node[right] at (3,3) {\scalebox{.8}{$=S_- $\;\;, }}; 
              \draw [-stealth,red] (2.3,2.5) arc (-30:30:10mm);
        
              \filldraw[fill=black, draw=black]  (5,3) circle (0.05 cm);
                 \draw [-stealth] (5,3) - - (6,3);
            \draw (6,3) - - (7,3);
          \node[right] at (7,3) {\scalebox{.8}{$=S_+ $}}; 
          \draw [-stealth,red] (6.3,2.5) arc (-30:30:10mm);

   \filldraw[fill=black, draw=black]  (1,1.5) circle (0.05 cm);
      \draw [stealth-] (2,1.5) - - (3,1.5);
            \draw (1,1.5) - - (2,1.5);
          \node[right] at (3,1.5) {\scalebox{.8}{$=S_-^{-1}$\;\;, }}; 
              \draw [stealth-,red] (2.3,1) arc (-30:30:10mm);
       
              \filldraw[fill=black, draw=black]  (5,1.5) circle (0.05 cm);
                 \draw [-stealth] (5,1.5) - - (6,1.5);
            \draw (6,1.5) - - (7,1.5);
          \node[right] at (7,1.5) {\scalebox{.8}{$=S_+^{-1}$}}; 
          \draw [stealth-,red] (6.3,1) arc (-30:30:10mm);
          
          \draw[snake=snake,segment length=5pt] (9,1.5)-- (11,1.5); 
              \filldraw[fill=black, draw=black]  (9,1.5) circle (0.05 cm);
 
          \node[right] at (11,1.5) {\scalebox{.8}{$=B^{-1}$}}; 
          \draw [stealth-,red] (10.3,1) arc (-30:30:10mm);
 
                 \filldraw[fill=black, draw=black]  (9,3) circle (0.05 cm);
              \draw[snake=snake,segment length=5pt] (9,3)-- (11,3); 
          \node[right] at (11,3) {\scalebox{.8}{$=B$}}; 
          \draw [-stealth,red] (10.3,2.5) arc (-30:30:10mm);

 \end{tikzpicture}}}
    \caption{Local connection matrices as we pass an oriented Stokes line (straight lines with arrows) or a branch-cut (wavy lines) for  a simple turning point.}
	\label{fig:Airy2}
\end{figure}
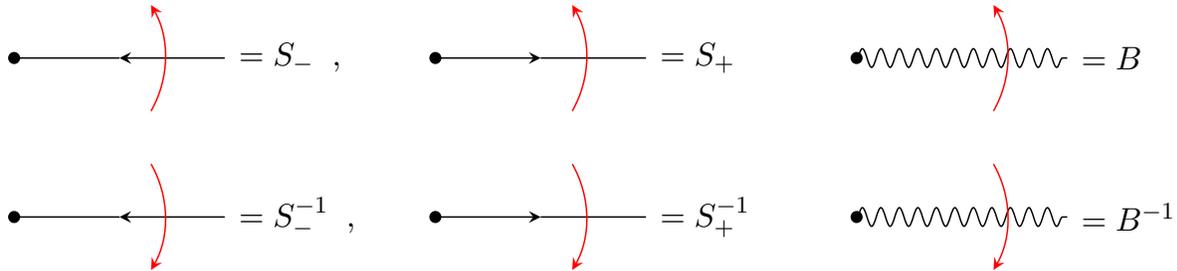

Whenever we cross a Stokes line, either $\psi_+$ or $\psi_-$ undergoes a jump.
The connection matrices for the wave-functions were found by Voros for a pure quartic potential \cite{voros-quartic}. His analysis has been further generalized and formalized in \cite{reshyper,dpham,ddpham,AKT1}. 
For simple turning points the full connection matrices can be split into ``local" connection matrices which encode the Stokes automorphisms around a given simple turning point 
and the Voros symbols \eqref{eq:VorosDef}, which connect wave functions in ``distant" regions, associated to different turning points.
The local connection matrices around a simple turning point coincide with those given by the Airy differential equation, \eqref{ODGp1} with $q=1$.
See appendix \ref{app:connmat} for the explicit computation.  In the Airy case, the complex $z$-plane splits in three equal wedges, as depicted in figure \ref{fig:Airy}. 
An arrow entering (exiting) the turning point corresponds to Stokes lines where $\psi_+$ ($\psi_-$)  jumps. 
It is also convenient not to talk about the jump of the solutions $\psi_\pm$ but of the jump of their
coefficients. Namely we write 
\be
\psi^{\text I}  = c_+^{\text I} \psi_+ + c_-^{\text I} \psi_- \,,\qquad \qquad
\psi^{\rm II}  = c_+^{\text{\rm II}} \psi_+ + c_-^{\text{\rm II}} \psi_-\,,
\ee
where $\psi_\pm$ are the wave-functions in a chosen reference wedge.\footnote{Unless specified otherwise, we take this reference wedge to be the one containing the asymptotic positive real axis, or its 
upper part, if the latter is on a Stokes line.} 
The local connection matrices as we pass a Stokes line or a branch-cut can be expressed as $2\times 2$ matrices acting on the coefficients $(c_+^{\rm N} , c_-^{\rm N} )$ 
as follows 
\be
c^{\rm II}  = A c^{\rm I}\,,  \qquad c^{\rm N} \equiv   \left(\begin{array}{c} c_+^{\rm N}  \\ c_-^{\rm N} \end{array}\right)\,,
\label{eq:cDef}
\ee
where $A$ is any of the $2\times 2$ matrices below (see figure \ref{fig:Airy2}):
\be
S_+  =  \left(\begin{array}{cc}
1 & i  \\
0 & 1
\end{array}\right) \,, \qquad 
S_-= \left(\begin{array}{cc}
1 & 0  \\
i & 1
\end{array}\right) \,, \qquad 
B =\left(\begin{array}{cc}
0 & -i  \\
-i & 0
\end{array}\right) \,.
\label{eq:monodromy}
\ee
The relations \eqref{eq:monodromy} apply locally around an arbitrary simple turning point. 

Voros symbols connect wave functions associated to different turning points.
If $A$ and $B$ are two simple turning points, from \eqref{eq:VorosDef2} we have 
\be
\psi_\pm^{(B)}(z) = \psi_\pm^{(A)}(z) a_{AB}^{\mp \frac 12}  \,,
\label{eq:psiasAB}
\ee
where $\psi_\pm^{(A,B)}$ are the Borel resummations (in the appropriate wedge of the complex plane) of the formal wave functions \eqref{eq:psiasPeven} normalized at $z_0=A,B$.
It is useful to define the Voros connection matrix $\mathcal{V}_\gamma$ as
\be
\mathcal{V}_\gamma = \left(\begin{array}{cc}
a_\gamma^{\frac 12}  & 0  \\
0 & a_\gamma^{-\frac 12} 
\end{array}\right) \,.
\label{eq:VorosVDef}
\ee 
A similar relation is used to define the corresponding formal power series $\widetilde{\mathcal{V}}_\gamma$.

\subsection{Rules to determine EQCs}
\label{subsec:EQCs}

Exact quantization conditions are 
conditions that we impose on a combination of $\psi_\pm$ in order to have a well-defined physical system. For bounded polynomial potentials with a stable discrete spectrum,
the case considered in this paper, EQCs are determined by demanding that the wave functions are square integrable over the real axis. 

Given a potential $Q(z)$ with only simple turning points, we start by determining all the Stokes lines using the definition \eqref{eq:StokesLineDef}.
Possible singular Stokes lines are avoided by assigning a phase to $\hbar$. The determination of the exact wave functions $\psi_\pm(z)$, Borel resummation
of the formal series \eqref{eq:psiasPeven}, follow from the matrices \eqref{eq:monodromy} and \eqref{eq:VorosVDef}. 
Let us denote by $\psi^{(\text P)} = c_+^{(\text P)} \psi_++c_-^{(\text P)} \psi_-$ and $\psi^{(\text N)} = c_+^{(\text N)} \psi_++c_-^{(\text N)} \psi_-$ the wave functions in the wedges containing the positive (P) and negative (N) real axis. If any of them is a Stokes line, we can infinitesimally shift the point above or below in the $z$-plane. The ending result does not depend on which shift we choose to make.
We define $c^{(\text N)} = {\cal M} c^{(\text P)}$, where
\be
{\cal M} =  \left(\begin{array}{cc}
m_{11} & m_{12}  \\
m_{21}& m_{22} 
\end{array}\right) 
\ee
is the total monodromy matrix along a given path connecting a point in the asymptotic positive real axis and a point
 in the negative asymptotic real axis. 
For real potentials square integrability requires that $c_+^{(\text P)} = 0$, which determines the EQCs.
Depending on the potential and in which Riemann sheet we end up when moving along the path, the requirement is  
\be
m_{12} = 0 \qquad \text{or}  \qquad m_{22} = 0 \,.
\ee
Since the coefficients $m_{ij}$ are products of the  matrices \eqref{eq:monodromy} and \eqref{eq:VorosVDef}, we see that eventually EQCs depend on Voros symbols only.
A good sanity check for the correctness of the procedure is to consider a closed path in the $z$-plane, in which case the connection matrix should reduce to the identity, 
given that wave functions are single-valued.

\begin{figure}
	\centering
	\includegraphics[width=0.25\linewidth]{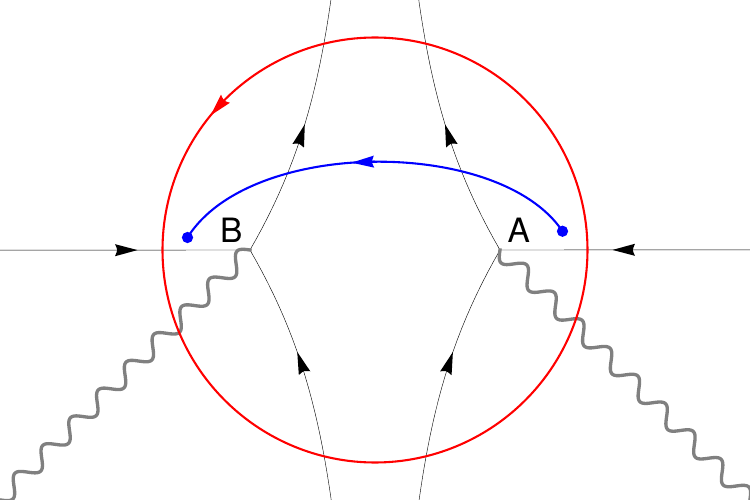} \qquad
		\includegraphics[width=0.25\linewidth]{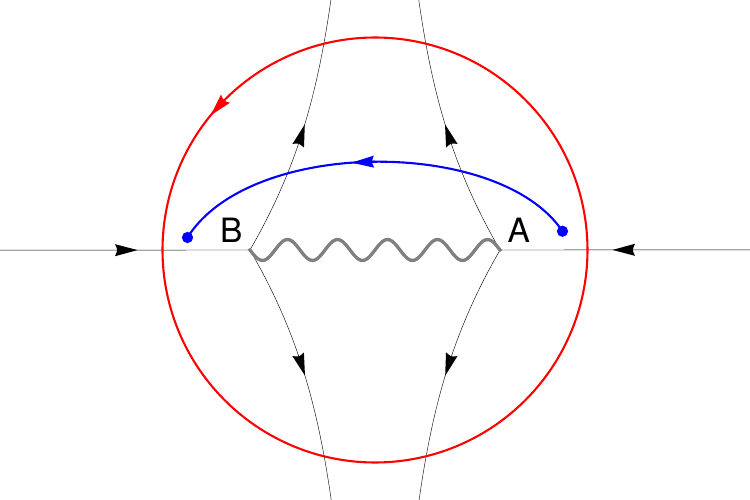} \qquad
			\includegraphics[width=0.25\linewidth]{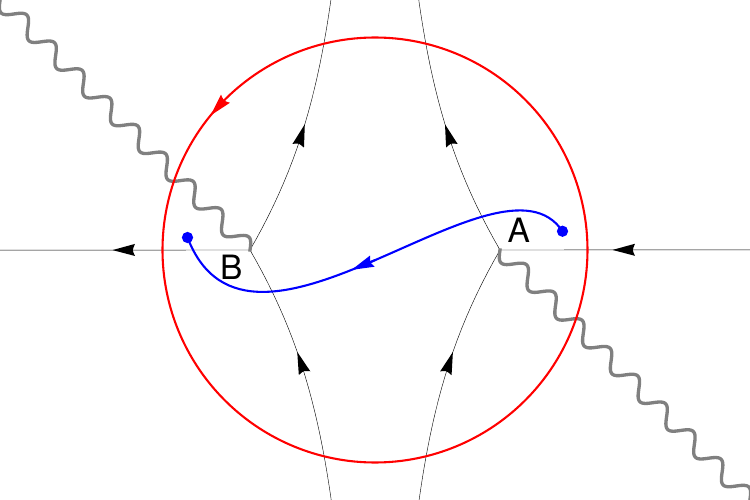} \\ 
			\vskip 7pt
			\small (a) \hspace{4.2cm} (b) \hspace{4.2cm} (c)
	\caption{Three arbitrary choices of branch-cuts and connecting paths (blue lines) to determine EQCs in the harmonic oscillator. The total monodromy around the red circles are given in \eqref{eq:monoHO}, the EQCs in \eqref{eq:EQCHO}. }
	\label{fig:HO}
\end{figure}

It is useful to consider a simple example, the harmonic oscillator. Setting $m=\omega=1$, we have
\be
Q_0 = z^2-2E\,,
\label{eq:Q0HO}
\ee
where $E\geq 0$. For $E\neq 0$, we get two turning points at $A = \sqrt{2E}$ and $B=-\sqrt{2E}$.  
We report in fig.\ref{fig:HO} the decomposition of the $z$-plane in the different Stokes regions delimited by regular Stokes lines.
Three possible choices of branch-cuts (wavy lines) and connection path (blue line) are shown.

As a first check, let us verify that the total monodromy is trivial. Starting from the region above the positive real axis in figure \ref{fig:HO}, we verify in the three cases (to be read from right to left as we move along the arrow) 
\begin{align}
(a): &  \qquad S_- B S_- \mathcal{V}_\gamma^{-1} S_- B S_- S_+ \mathcal{V}_\gamma S_+ = I\,, \nn \\
(b): & \qquad \quad B^2 S_- S_+ \mathcal{V}_\gamma S_+ S_- S_+ \mathcal{V}_\gamma S_+ = I\,,\label{eq:monoHO}  \\
(c): & \qquad S_-  B S_- \mathcal{V}_\gamma^{-1} S_- S_+ B  S_+ \mathcal{V}_\gamma S_+ = I\,. \nn
\end{align}
We determine the EQC associated to the three blue paths in figure \ref{fig:HO}. In the region slightly above the positive real axis the normalizable wave function is $\psi_-$. So, we start with $c_+^{\text{P}}=0$ and proceed along the blue curve towards the negative real axis, slightly above or below it. Demanding convergence of the wave function requires that $c_+^{\text{N}}=0$ in cases (a) and (b), while in case (c) we require  $c_-^{\text{N}}=0$. We have
\begin{align}
(a): & \qquad (S_+ \mathcal{V}_\gamma S_+)_{12} = 0 \,, \nn \\
(b): & \qquad (S_+ \mathcal{V}_\gamma S_+)_{12} = 0  \,, \label{eq:EQCHO} \\
(c): & \qquad (S_-^{-1} \mathcal{V}_\gamma S_+)_{22} = 0\,. \nn
\end{align}
The EQCs above are all equivalent to 
\be
a_\gamma = -1\,.
\label{eq:CFHO}
\ee
At a first sight it seems that the computation of $a_\gamma$ requires to consider the full expansion of $P_\text{even}$ in $\hbar$, but, luckily enough, 
all terms but one vanish (after all, this is expected since the harmonic oscillator is exactly solvable!).
A simple way to show this is to blow up the contour $\gamma$ to a parametrically large circle in the $z$-plane. Since $P_0 \propto z$, we see from the recursion relation \eqref{eq:recrel}
and the form of the first coefficients \eqref{eq:Pcoeff} that $P_{2n}\propto z^{1-4n}$ for large $z$. Hence all contributions, but $P_0$, vanish. The integral over $P_0$ is elementary and gives the correct energy eigenvalues:
\be
a_\gamma = e^{\frac{2i \pi E}{\hbar}}\qquad \Longrightarrow E_n = \hbar \Big(n+\frac 12\Big)\,, \quad n\in \mathbb{N}\,.
\ee

A few comments are in order. 
\begin{enumerate}
\item The wave-function that jumps over the positive real axis is $\psi_+$ and this fixes the arrow in the Stokes line as in the Airy case.
All other arrows follow. Two nearby Stokes lines emanating from a simple turning point have equal (opposite) orientation if a branch-cut (does not) divides them.
Asymptotic Stokes lines in the same direction must have the same orientation.
\item  Branch-cuts can be inserted between turning points or go at infinity. As far as periods are concerned both options are equivalent, since the branch-cuts are of square root type. For wave functions, however, this is not the case, because of the extra factor $\widetilde P^{-1/2}_{\text{even}}$ in \eqref{eq:psiasPeven}, which is not automatically taken into account when the branch-cuts are inserted between turning points. In the harmonic oscillator, $P^{-1/2}_{\text{even}}\sim z^{-1/2}$ for large $z$,
and hence we have a total monodromy around infinity given by $B^2 = -I$. This explains the origin of the $B^2$ factor in case (b) of \eqref{eq:monoHO}.
\item The (asymptotic) positive and negative real axis are Stokes lines. The orientation of the Stokes line in the negative real axis determines the EQC.
We have $c_{12}=0$ or $c_{22}=0$ if $\psi_+$ or $\psi_-$ respectively jump over the real negative axis. The same applies for more general potentials with real parameters for which the real axis is on Stokes lines.
\item  Whenever the cycle between $A$ and $B$ does not have a branch-cut, the integral from $A$ to $B$ is opposite to the one from $B$ to $A$,
while it is equal if there is a branch-cut and the path circles through the opposite side of it. That is why in \eqref{eq:monoHO} we have $\mathcal{V}_\gamma$ and $\mathcal{V}_\gamma^{-1}$ in (a) and (c), but only $\mathcal{V}_\gamma$ in (b).
\end{enumerate}
We also have
\be
S_\pm B = B S_\mp\,, \qquad B \mathcal{V}_\gamma = \mathcal{V}_\gamma^{-1} B\,.
\label{eq:SBRel}
\ee
We will make extensive use of these rules to derive EQCs in the next sections.

As anticipated in the introduction, there are at least two ways to make use of EQCs in exact WKB to determine the energy spectrum $E_n$ of the system:
\begin{enumerate}
\item We can directly determine $E_n$ as those values of $E$ for which \eqref{eq:Fexsol} is satisfied.
In this approach $E$ is taken to be of ${\cal O}(1)$, and $a_{\gamma_i}$ are determined from \eqref{eq:qPeriod} and \eqref{eq:VorosDef}.
Typically such $E_n$ are found numerically.
\item We can consider the ``downgraded" form 
\be
F[\widetilde a_{\gamma_i}(E) ] = 0\,,
\label{eq:Fasymsol}
\ee
where we undo the Borel resummation and Voros symbols in \eqref{eq:Fexsol} are replaced by their formal power series.
In \eqref{eq:Fasymsol} we then replace $E\rightarrow \widetilde E_n$, where $\widetilde E_n$ are in general transseries which are obtained by demanding 
\eqref{eq:Fasymsol} order by order in $\hbar$ and $\exp(-1/\hbar)$. 
The final energy spectrum is obtained by an appropriate Borel resummation of $\widetilde E_n$: $E_n = s(\widetilde E_n)$.
 \end{enumerate}
Since we are interested in the series expansion of energy eigenvalues, approach 2. is the one which will be mostly considered in this paper. However it is worth noting that there are systems, such as the anharmonic oscillator with no mass term, whose spectrum does not possess a transseries representation in $\hbar$, which makes approach 2. fail.\footnote{By this we mean within ordinary perturbation theory. As we will see, approach 2 within EPT works also in these cases.} Appendix~\ref{app:purequartic} studies such a system in detail.

\section{Anharmonic oscillators}
\label{sec:pure-anharmonic}

In this section we analyze EQCs for anharmonic oscillators in some detail. We mostly focus on the quartic case, and then generalize to higher order anharmonic oscillators.
We show that, at fixed moduli of the quartic anahamornic potential, there exist eight EQCs in different wedges in the $\hbar$ complex plane.\footnote{Among other things, a classification of the different EQCs for the quartic oscillator already appeared in \cite{DELABAERE1997180} (see in particular fig.5 there). The perspective in \cite{DELABAERE1997180} was however a bit different as  they were mostly interested on the ramifications of the energy eigenvalues as the (complex) moduli of the potential are varied. Here the moduli are held fixed and we classify
the regions depending on the values of the energy and the phase of $\hbar$. The two analysis are however not totally independent.}
They include a ``sweet spot'' region where the EQC is particularly simple.
We show how the Borel summability of the asymptotic series associated to the energy eigenvalues $\widetilde E_n$ follows from the existence of this ``sweet spot''
region in the limit where two simple turning points collapse to a double turning point. 
More specifically, we show how Borel summability of $\widetilde E_n$ can be established without making direct use of connection matrices for double turning points,
as done in \cite{ddpham}. We then generalize our findings to higher order potentials.

\subsection{The quartic anharmonic potential in the $\hbar$ complex plane}
\label{sec:region-anharm}

\begin{figure}
\centering
\begin{tabular}{lccr}
\begin{subfigure}{0.22\textwidth}
\includegraphics[width=\textwidth]{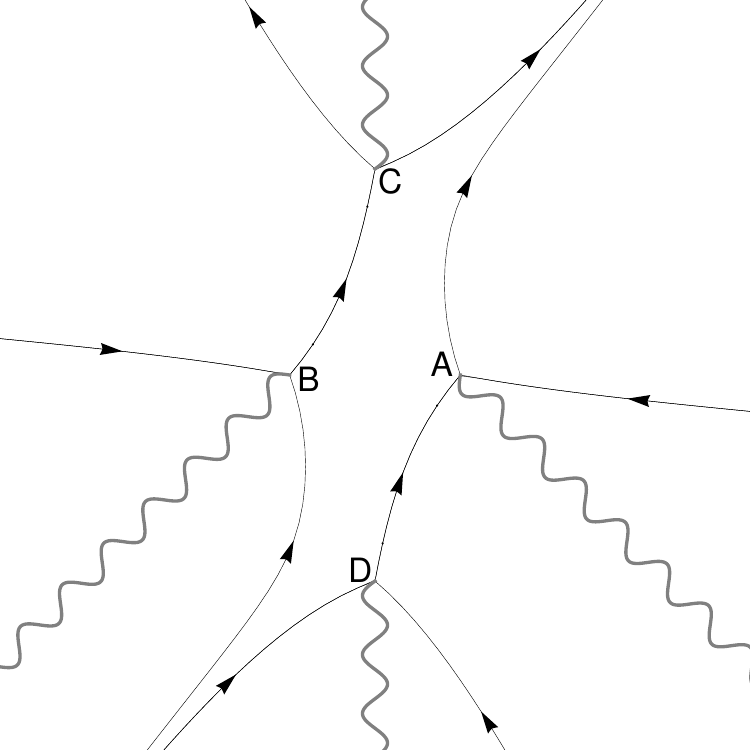}
\caption{\centering $\arg\hbar = -\alpha$}
\end{subfigure}
&
\begin{subfigure}{0.22\textwidth}
\includegraphics[width=\textwidth]{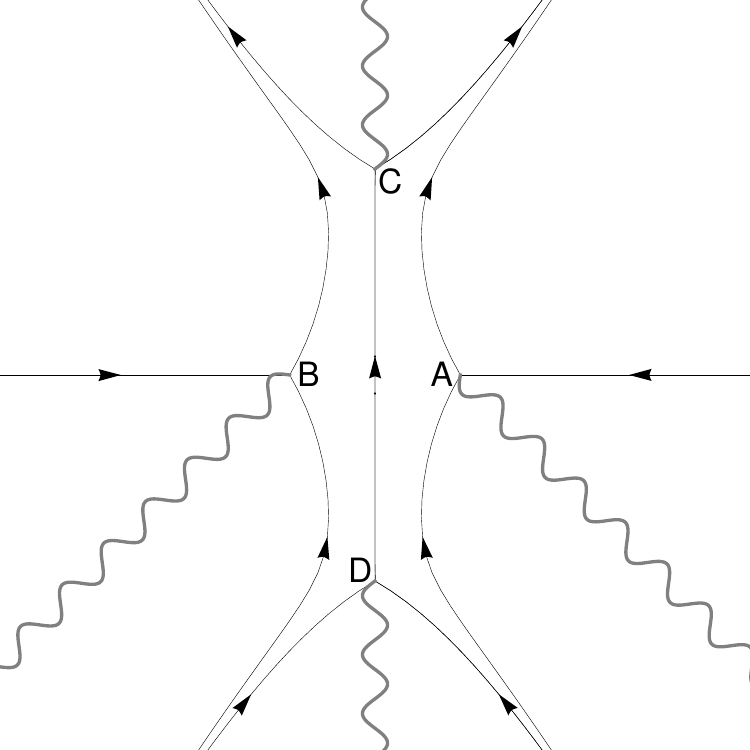}
\caption{\centering $\arg\hbar = 0$}
\end{subfigure}
&
\begin{subfigure}{0.22\textwidth}
\includegraphics[width=\textwidth]{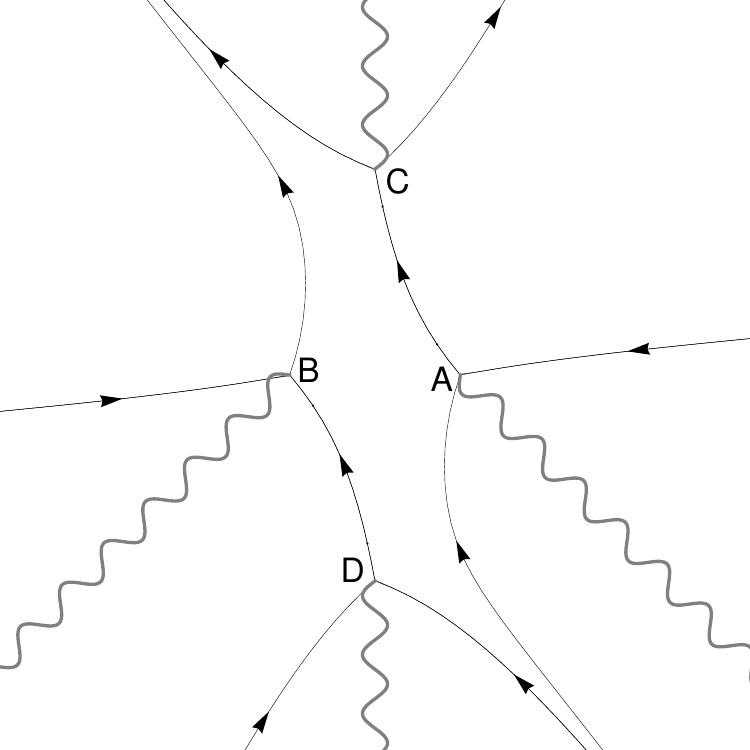}
\caption{\centering $\arg\hbar = \alpha$}
\end{subfigure}
&
\begin{subfigure}{0.22\textwidth}
\includegraphics[width=\textwidth]{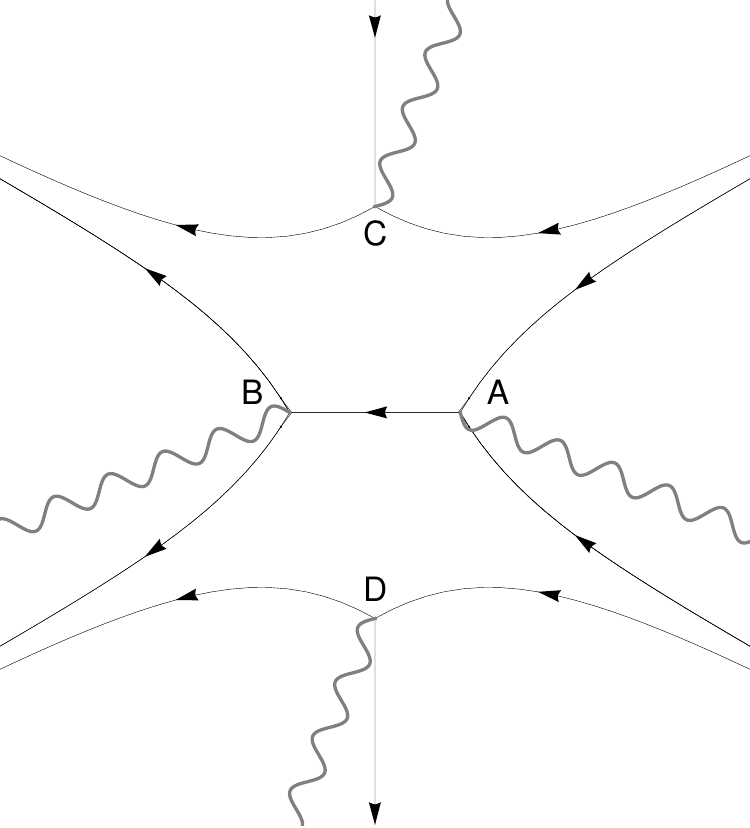}
\caption{\centering $\arg\hbar = \pi/2$}
\end{subfigure}
\end{tabular}
\caption{Phases of $\hbar$ with $-\pi/2< \text{arg}\,\hbar\leq\pi/2$ for which critical Stokes lines appear in the anharmonic oscillator \eqref{eq:PotAnh} with $E>0$.}
\label{fig-anharm-crit}
\end{figure}

We fix the moduli of the quartic anharmonic potential and take in \eqref{eq:Qdef}
\begin{equation}
V(z) = \frac{1}{2}\left(z^2+z^4\right)\,,
\label{eq:PotAnh}
\end{equation}
and $E$ real and positive. For any $E>0$ we have four simple turning points, two reals and two purely imaginary.
We report in figure \ref{fig-anharm-crit} the four phases of $-\pi/2 < \text{arg}\,\hbar\leq \pi/2$ for which singular Stokes lines appear.
Four mirror phases are present for $\pi/2 \leq |\text{arg}\,\hbar|\leq \pi$, as given by \eqref{eq:StokesInv}.
The turning points are labelled as in figure \ref{fig-anharm-crit}.
We have the ``perturbative period'' $a_{AB}$, the ``non-perturbative period'' $a_{DC}$, and four ``diagonal" periods. 
Putting our branch cuts from the turning point to infinity as in figure \ref{fig-anharm-crit}, we have 
\begin{equation}
\begin{aligned}
\arg \Pi_{BA} &= \frac{\pi}{2},\quad
\arg \Pi_{CD} = 0,\\ 
\arg \Pi_{CA} &= \arg \Pi_{BD} = -\arg \Pi_{AD} =  -\arg \Pi_{CB} = \alpha(E)\,.
\end{aligned}
\end{equation}
The angle $\alpha(E)$ will play an important role in our analysis. We do not report its full analytic expression, which is complicated.
As far as our analysis is concerned, what really matters is its expansion for small $E>0$, which reads
\begin{equation}
\alpha(E) = \arctan\left( \frac{ \Pi_{0,BA}(E)}{i \Pi_{0,CD}(E)}\right) \approx \frac{3 \pi}{8} E +O\left(E^2\right).
\label{eq:defalpha}
\end{equation}
The key feature, which crucially relies on the mass term in the potential, is that $\alpha(E\rightarrow 0)= 0$.

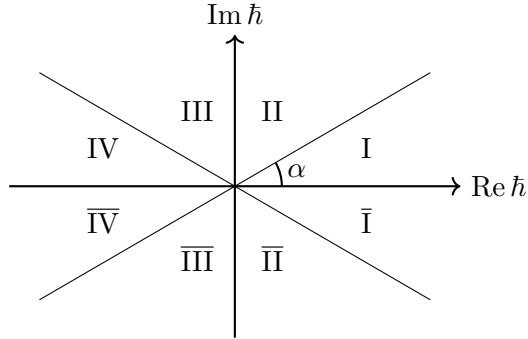
\begin{figure}
\begin{center}
\begin{tikzpicture}[scale=1]
\draw[->, thick] (-3,0)--(3,0) node[right]{$\textrm{Re}\,\hbar$};
\draw[->, thick] (0,-2)--(0,2) node[above]{$\textrm{Im}\,\hbar$};
\begin{scope}[rotate=30]
\draw[black] (-3,0) -- (3,0);
\end{scope}
\begin{scope}[rotate=-30]
\draw[black] (-3,0) -- (3,0);
\end{scope}
\draw[black, thick] (0.625,0) arc (0:30:.625);
\draw[] (.825,.2) node[]{$\alpha$};
\draw[] (1.75,.5) node[]{$\rm I$};
\draw[] (1.75,-.45) node[]{$\overline{\text{\rm I}}$};
\draw[] (.5,1) node[]{$\rm II$};
\draw[] (.5,-.95) node[]{$\overline{\text{\rm II}}$};
\draw[] (-1.75,.5) node[]{$\text{IV}$};
\draw[] (-1.75,-.45) node[]{$\overline{\text{IV}}$};
\draw[] (-.5,1) node[]{$\text{III}$};
\draw[] (-.5,-.95) node[]{$\overline{\text{III}}$};
\end{tikzpicture}
\end{center}
\caption{The eight regions delimited by singular Stokes lines in the complex $\hbar$ plane for $E>0$.}
\label{fig-regs}
\end{figure}

The eight singular Stokes line configurations define corresponding eight wedges, or regions, in the complex $\hbar$ plane, reported in  figure \ref{fig-regs}.
In each wedge, we can vary freely the argument of $\hbar$ without any Stokes jump. In the upper half plane we label
\begin{itemize}
\item Region $\rm I$: $0 <  \arg \hbar <  \alpha(E)$,
\item Region $\rm II$: $\alpha(E)<  \arg \hbar < \frac{\pi}{2} $,
\item Region $\text{III}$: $\frac{\pi}{2} < \arg \hbar < \pi-\alpha(E) $,
\item Region $\text{IV}$: $\pi-\alpha(E) < \arg \hbar < \pi$.
\end{itemize}
The reflections of these regions in the lower half plane are labeled by bars, as in figure \ref{fig-regs}.
It is useful to report the Stokes jumps of some relevant periods, derived using the DDP formula \eqref{eq:DDP}. The perturbative cycle is subject to the following jumps:
\begin{equation}
\begin{aligned} 
a^{\rm II}_{AB} &= (1+a_{AC}^{\rm I})^{-1}(1+a_{DB}^{\rm I})^{-1}a^{\rm I}_{AB}\,, \\
a^{\rm I}_{AB} &= (1+a_{DC}^{\overline{\text{\rm I}}})^{-2}a^{\overline{\text{\rm I}}}_{AB} \,, \\
a^{\overline{\text{\rm I}}}_{AB} &= (1+a_{DA}^{\overline{\text{\rm II}}})^{-1}(1+a_{BC}^{\overline{\text{\rm II}}})^{-1} a^{\overline{\text{\rm II}}}_{AB}\,.
\label{ddp-pert}
\end{aligned}
\end{equation}
We also have that
\begin{equation}
\begin{aligned} 
 a^{\rm I}_{DB} &= (1+a_{DC}^{\overline{\text{\rm I}}})^{-1}a^{\overline{\text{\rm I}}}_{DB},\\
a^{\rm I}_{AC} &= (1+a_{DC}^{\overline{\text{\rm I}}})^{-1}a^{\overline{\text{\rm I}}}_{AC},\\
\end{aligned}
\label{ddp-diag-zero}
\end{equation}
and
\begin{equation}
a_{DC}^{\text{III}}=(1+a_{AB}^{\rm II})^2a_{DC}^{\rm II}.
\label{ddp-DC}
\end{equation}
The Stokes jumps for other cycles can be obtained from the ones above by composition of periods and geometric properties of intersection numbers.
Stokes jumps occurring in the regions with $\text{Re}\,\hbar<0$ are determined using \eqref{eq:StokesInv}. For example, from the first relation in \eqref{ddp-pert}
we get
\be
a^{\overline{\text{\rm III}}}_{AB} = (1+a_{CA}^{\overline{\text{\rm IV}}})(1+a_{BD}^{\overline{\text{\rm IV}}}) a^{\overline{\text{\rm IV}}}_{AB}\,,
\ee
and similarly for the other cases.

\subsection{Exact quantization conditions}

\begin{figure}
\centering
\begin{tabular}{lccr}
\begin{subfigure}{0.22\textwidth}
\includegraphics[width=\textwidth]{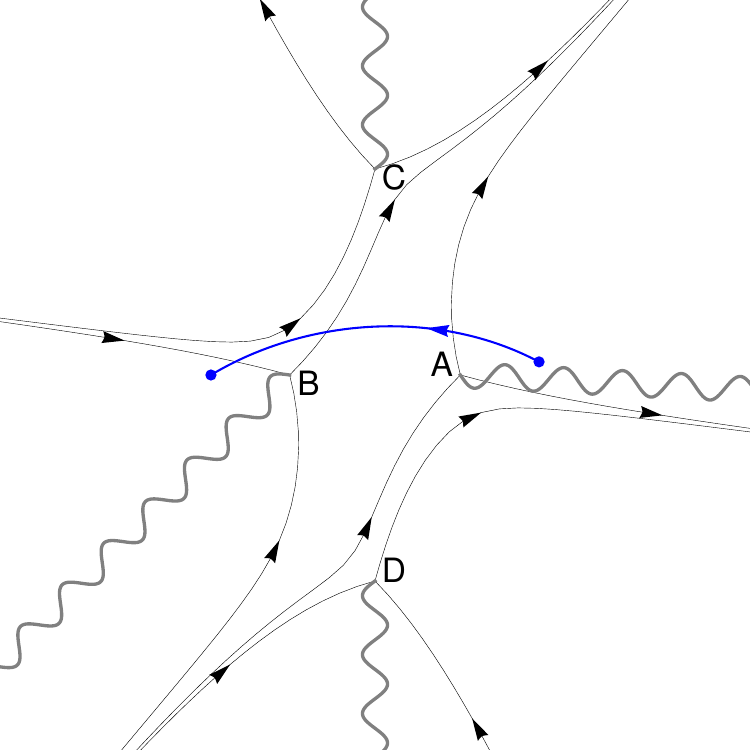}
\caption{Region $\overline{\text{\rm II}}$}
\end{subfigure}
&
\begin{subfigure}{0.22\textwidth}
\includegraphics[width=\textwidth]{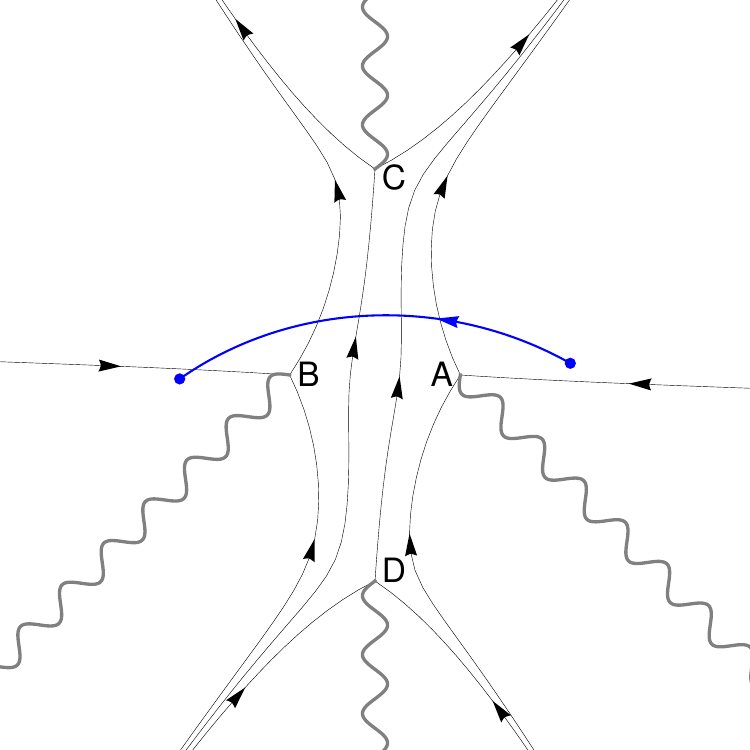}
\caption{Region $\overline{\text{\rm I}}$}
\end{subfigure}
&
\begin{subfigure}{0.22\textwidth}
\includegraphics[width=\textwidth]{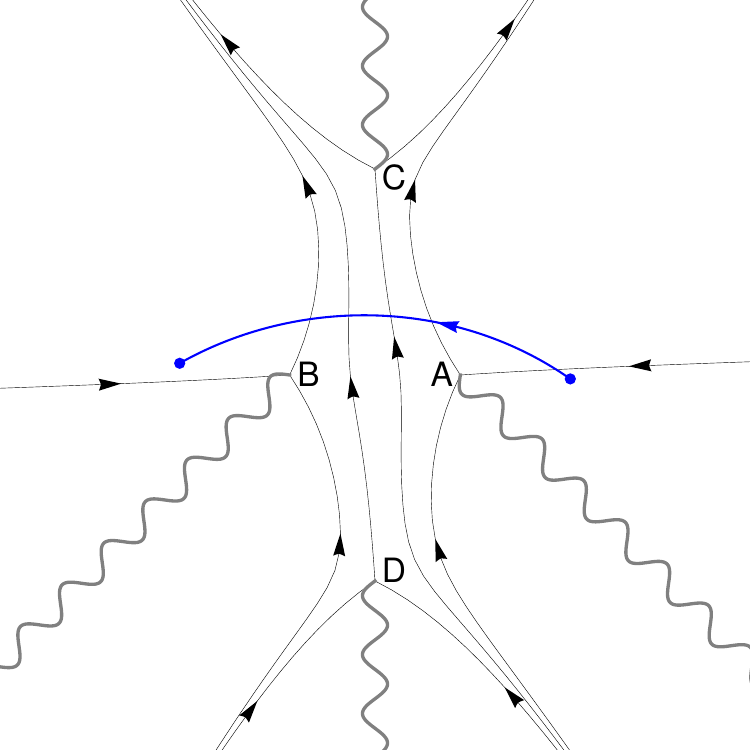}
\caption{Region I}
\end{subfigure}
&
\begin{subfigure}{0.22\textwidth}
\includegraphics[width=\textwidth]{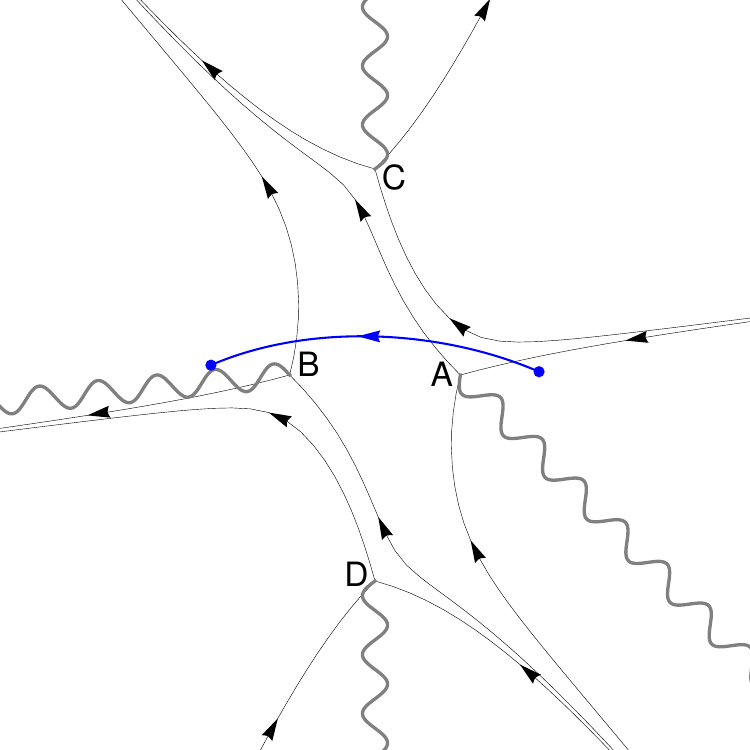}
\caption{Region II}
\end{subfigure}
\end{tabular}
\caption{EQCs and connection paths (blue lines) for the anharmonic oscillator \eqref{eq:PotAnh} with $E>0$ for the four regions with $\text{Re}\,\hbar >0$.}
\label{fig-anharm-eqc}
\end{figure}

We can now use the techniques of section \ref{sec:basics} to find the EQCs for the anharmonic oscillator.
The associated connection paths in the $z$ plane are reported as blue lines in figure \ref{fig-anharm-eqc} in the various regions.
In regions $\rm II$ and $\overline{\text{\rm II}}$ the EQCs are quite simple.
We have
\begin{equation}
\begin{aligned}
\left(S_+\mathcal{V}_{AB}S_+S_-\right)_{12}&=0\Rightarrow a^{\rm II}_{AB}+1 =0,\\
\left(S_-S_+\mathcal{V}_{AB}S_+\right)_{12}&=0\Rightarrow a^{\overline{\rm II}}_{AB}+1 =0.
\end{aligned}
\label{simple-eqc}
\end{equation}
This is similar to the purely perturbative quantization condition of the harmonic oscillator. 
In regions $\rm I$ and $\overline{\text{\rm I}}$ we obtain more complicated expressions, namely
\begin{equation}
\begin{split}
\left(S_+\mathcal{V}_{DB}S_+\mathcal{V}_{CD}S_-^{-1}\mathcal{V}_{AC}S_+\right)_{12} & =0
\Rightarrow 1+a^{\rm I}_{AC}+a^{\rm I}_{DB}+a^{\rm I}_{AC}(1+a^{\rm I}_{CD})a^{\rm I}_{DB}=0\,,  \\
\left(S_+\mathcal{V}_{CB}S_-^{-1}\mathcal{V}_{DC}S_+\mathcal{V}_{AD}S_+\right)_{12}& =0
\Rightarrow  1+(1+a^{\overline{\text{\rm I}}}_{AD})(1+a^{\overline{\text{\rm I}}}_{CB})a^{\overline{\text{\rm I}}}_{DC}=0\,. 
\end{split}
\label{eq:EQCIIb}
\end{equation}
By geometrically composing the periods we can rewrite \eqref{eq:EQCIIb} as
\begin{equation}
\begin{aligned} 
1+(1+a^{\rm I}_{AC})^{-1}(1+a^{\rm I}_{DB})^{-1}a^{\rm I}_{AB} &=0, \\
1+(1+a^{\overline{\text{\rm I}}}_{DA})(1+a^{\overline{\text{\rm I}}}_{BC})a^{\overline{\text{\rm I}}}_{AB}&=0\,. 
\label{eqcs-reduced}
\end{aligned}
\end{equation}
Alternatively, the EQCs can be written in terms of a choice of only two periods, though this results in a less compact form. In the form \eqref{eqcs-reduced} it is manifest that the EQCs in region I/$\overline{\rm I}$ are compatible with those in region II/$\overline{\rm II}$ as long as one accounts for the Stokes jumps listed in \eqref{ddp-pert}. On can further check that they are also compatible with each other through \eqref{ddp-diag-zero}. 

We then note that the wedges II and $\overline{\rm II}$ are ``sweet spot'' regions where we get a simple EQC involving only the perturbative Voros symbol. 
This region is somehow hidden by the fact that, for any finite $E>0$, the limit $\text{arg}\, \hbar \rightarrow 0$ leads us to region I.
The EQCs in this region include non-perturbative corrections, which might at first seem puzzling, due to the long known Borel summability 
of the $\hbar$ expansion of the anharmonic potential \cite{Graffi:1970erh}. If we determine the spectrum using EWKB as discussed in point 1. below \eqref{eq:SBRel}, 
the exact spectrum is in fact determined by \eqref{eqcs-reduced} where the non-perturbative cycles should be included. 

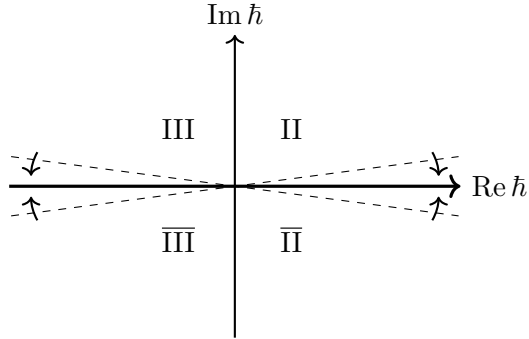
\begin{figure}[t!]
\begin{center}
\begin{tikzpicture}[scale=1]
\draw[->, very thick] (-3,0)--(3,0) node[right]{$\textrm{Re}\,\hbar$};
\draw[->, thick] (0,-2)--(0,2) node[above]{$\textrm{Im}\,\hbar$};
\begin{scope}[rotate=7.5]
\draw[black,dashed] (-3,0) -- (3,0);
\end{scope}
\begin{scope}[rotate=-7.5]
\draw[black,dashed] (-3,0) -- (3,0);
\end{scope}
\draw[black, thick, ->] (2.625,0.45) arc (30:0:.625);
\draw[black, thick, ->] (2.625,-0.45) arc (-30:0:.625);
\draw[black, thick, ->] (-2.625,0.45) arc (150:180:.625);
\draw[black, thick, ->] (-2.625,-0.45) arc (210:180:.625);
\draw[] (.75,.75) node[]{$\text{\rm II}$};
\draw[] (.75,-.7) node[]{$\overline{\text{\rm II}}$};
\draw[] (-.75,.75) node[]{$\text{III}$};
\draw[] (-.75,-.7) node[]{$\overline{\text{III}}$};
\end{tikzpicture}
\end{center}
\caption{The regions of the complex $\hbar$ plane in the limit $e_0\rightarrow 0$.}
\label{fig-vanish}
\end{figure}

On the other hand, if we want to relate EQCs with the asymptotic series of energy eigenvalues, we should consider the approach in point 2. below \eqref{eq:SBRel}.
We undo the Borel resummation implicit in the above EQCs, and turn them into formal power series equations, with 
\begin{equation}
E\rightarrow \widetilde E = \widetilde E_{\text{P}} + \widetilde E_{\text{NP}}\,, \qquad \widetilde E_{\text{P}} = e_0 + \sum_{k= 1}^\infty e_{k} \hbar^{k}
\label{eq:EtildeAN}
\end{equation}
and $\widetilde E_{\text{NP}}$ the non-perturbative transseries terms.
Independently of the region we start in, the first result one obtains when asymptotically solving the EQCs is $e_{0}=0$, i.e. $\widetilde E_{\text{P}} \sim O(\hbar)$. 
At fixed argument of $\hbar$, the limit $e_0\rightarrow 0$ forces us to be in regions ${\rm II}$/$\overline{\text{\rm II}}$/$\text{III}$/$\overline{\text{III}}$, respectively, since $\alpha(0)=0$.
See figure \ref{fig-vanish} for an illustration of the complex $\hbar$ plane in this limit. Thus, the EQCs with non-perturbative terms are never realized unless one takes some very unnatural limit. The absence of non-perturbative Voros symbols in the EQCs in the surviving regions implies that we can have solutions where  $\widetilde E_{\text{NP}} = 0$, i.e. perturbative
asymptotic series with no associated transseries. In fact, for $e_0\rightarrow 0$ the two turning points A and B collapse to a double turning point at $z_0=0$ and the EQC \eqref{simple-eqc} applies in the entire $\hbar$ complex plane with the exception of the real axis. However, as we will see, the EQCs derived before are {\it smooth} in the limit $e_0\rightarrow 0$ and there is no need to derive EQCs where double turning points are present to start with, as done in \cite{ddpham}.

We parametrise the Voros cycle $a_{AB}$ as 
\begin{equation}
{a}_{AB} \equiv e^{-2\pi i \left(t+\frac{1}{2}\right)}\,.
\end{equation}
The downgraded version of the EQC \eqref{simple-eqc} for $e_0\rightarrow 0$ reads then 
\be
\widetilde t(\widetilde E;\hbar)=n\,, \qquad n\in\mathbb{N}. 
\label{eq:EQCfort}
\ee
Equation \eqref{eq:EQCfort} allows us to determine the asymptotic series for the energy eigenvalues. 
Starting from finite $E$, we compute the period $\widetilde \Pi_{AB}$ perturbatively, and then replace $E$ as in \eqref{eq:EtildeAN}.
For the potential \eqref{eq:PotAnh} we get\footnote{As shown in \cite{ddpham}, by taking $e_0=0$ directly, 
the series \eqref{eq:ttildeAS} can also be determined by looking at the residue of the simple pole in $\widetilde P(z)$ at $z=0$.}
\be
\widetilde t =  \Big(e_1 - \frac 12\Big)  +  \hbar \Big(e_2 -\frac 3{16} (1+4 e_1^2)\Big)  + \hbar^2 \Big(e_3 -\frac 32 e_2 e_1 +\frac{5}{64} e_1(17+28e_1^2) \Big)
+ {\cal O}(\hbar^3)\,.
\label{eq:ttildeAS}
\ee
Demanding \eqref{eq:EQCfort} order by order in $\hbar$ fixes all energy coefficients $e_k$ and allows us to determine all the asymptotic series $\widetilde E_n$ at once for any $n$:
\be
\widetilde E_n =  \Big(n+\frac 12\Big) + \frac{3}{16} \bigg(1+4\Big(n+\frac 12\Big)^2\bigg)\hbar - \frac{1}{64}\Big(n+\frac 12\Big)\bigg(68\Big(n+\frac 12\Big)^2+67\bigg) \hbar^2 + {\cal O}(\hbar^3)\,.
\label{eq:expsolE}
\ee
Since only the perturbative cycle $\widetilde a_{AB}$ is involved, no non-perturbative terms appear. 
Proving the Borel summability of the series $\widetilde{E}_n$ requires a further analysis of the $\hbar\in \mathbb{R}^+$ case, which is discussed next.

\subsection{Borel summability}
\label{subsec:anharm-borel}

Borel summability of the energy eigenvalues $\widetilde{E}_n$ requires that 
\begin{equation}
s_{0+}(\widetilde{E}_n) = s_{0-}(\widetilde{E}_n)\,.
\end{equation}
By means of \eqref{app:StokesDef1} and \eqref{app:StokesDef2}, this is equivalent to show that 
\begin{equation}
\dot{\Delta}_A \widetilde E_n = 0, \quad \forall A \in \mathbb{R}^+ \,,
\label{alien-swap}
\end{equation}
where $\dot{\Delta}_A$ is the dotted alien derivative defined in \eqref{app:StokesDef2}.
Applying $\dot \Delta_A$ to \eqref{eq:EQCfort} and, using simple alien calculus rules,\footnote{For a physics oriented introduction to alien calculus see e.g. the pedagogical reviews \cite{Dorigoni:2014hea,abs}, while for a more mathematics oriented presentation see e.g. \cite{Sauzin}.}, we have \cite{ddpham}
\begin{equation}
0 = \dot \Delta_A \widetilde{t}(\widetilde{E}_n(\hbar);\hbar) = \dot{\Delta}_A \widetilde t(e_1;\hbar)|_{e_1=\widetilde{E}_n} + (\dot{\Delta}_A \widetilde{E}_n ) \,\partial_{e_1} \widetilde t(e_1;\hbar)\Big|_{e_1=\widetilde{E}_n} \,,
\label{Alien-calc-t-to-E}
\end{equation}
where $\widetilde t(e_1;\hbar)$ is the asymptotic series \eqref{eq:ttildeAS} where all $e_{k}$ but $e_1$ are set to zero. The condition $e_1= \widetilde E_n$
means replacing $e_1$ with the asymptotic series $\widetilde E_n$, solution of \eqref{eq:EQCfort}, whose first order terms are given in \eqref{eq:expsolE}.
As explained before, for $e_0\rightarrow 0$ regions II and $\overline{\rm II}$ cover the whole cut complex plane.
Since $\partial_{e_1} \widetilde t(e_1;\hbar)$ can never vanish, establishing \eqref{alien-swap} is equivalent to establish
$\dot \Delta_A \widetilde{t}=0$, namely 
\begin{equation}
(s_{0+}-s_{0-})\widetilde{t}(e_1;\hbar)|_{e_1=\widetilde{E}_n} = \lim_{e_0\rightarrow 0} (t^{\rm II}(e_0+\hbar e_1;\hbar)-t^{\overline{{\rm II}}}(e_0+\hbar e_1;\hbar))|_{e_1=\widetilde{E}_n} = 0
\,.
\label{eq:t-2-t-2bar}
\end{equation}
The relation between $t^{{\rm II}}$ and $t^{\overline{{\rm II}}}$ can be derived by combining the jumps at $\arg\hbar=-\alpha,0,\alpha$
between regions II and $\overline{\rm II}$ determined in the previous section. We get
\begin{equation}
a_{AB}^{\overline{\text{\rm II}}}=\left(1+a_{DA}^{\text{\rm II}}\left(1+a_{AB}^{\text{\rm II}}\right)\right)\left(1+a_{BC}^{\text{\rm II}}\left(1+a_{AB}^{\text{\rm II}}\right)\right)a_{AB}^{\text{\rm II}}.
\label{triple-jump}
\end{equation}
Taking the logarithm, we write
\begin{equation}
\lim_{e_0\rightarrow 0} \Big(t^{\rm II}-t^{\overline{\rm II}}\Big) = \lim_{e_0\rightarrow 0} 
\frac{1}{2\pi i}
\bigg\{
\log\Big(1+a_{DA}^{\rm II}\left(1+a_{AB}^{\rm II}\right)\Big)+
\log\Big(1+a_{BC}^{\rm II}\left(1+a_{AB}^{\rm II}\right)\Big)
\bigg\} \,,
\label{t-disc-noneven}
\end{equation}
Since the potential is even, $V(z) = V(-z)$, one can further simplify
\begin{equation}
\lim_{e_0\rightarrow 0} \Big(t^{\rm II}-t^{\overline{\rm II}}\Big) = \lim_{e_0\rightarrow 0} \frac{1}{\pi i}\log\left(1+\sqrt{\frac{a_{DC}^{\text{\rm II}}}{a_{AB}^{\text{\rm II}}}}\left(1+a_{AB}^{\text{\rm II}}\right)\right) \,.
\label{t-disc}
\end{equation}
When the EQC \eqref{simple-eqc} is imposed, the right hand side of \eqref{t-disc} vanishes unless $a_{DC}$ is singular. The regularity of $a_{DC}$, for sufficiently small $\hbar$, has been formally proven in \cite{dpham}. It is however useful to work out $\widetilde a_{DC}$ to one loop level. It is convenient to determine $a_{DC}$ in terms of the perturbative cycle $a_{AB}$, which amounts to changing variables from $E$ to $t$. 
To leading order, inverting the relation $E$ as a function of $t$ is merely
\begin{equation}
\widetilde{E}(t;\hbar)=\mu  \hbar + O\left(\hbar^2\right), \quad \mu \equiv t+\frac{1}{2}.
\label{E-RS}
\end{equation}
Assuming finite $E$, we calculate first the series for the non-perturbative period.\footnote{This series can be obtained to high order using, for example, the differential operator method introduced in \cite{huang-dif-op}, see e.g. appendix C of \cite{Emery:2020qqu} for a concrete implementation in a similar case.} Then we replace $E$ by \eqref{E-RS} to find
\begin{equation}
\begin{multlined}
\widetilde{\Pi}_{DC}\left(E(t;\hbar);\hbar\right) = 
\left( -\frac{4}{3}+2 \mu  \hbar  \log \left(\frac{  \hbar }{8}\right)+2 \mu  \hbar  \left(\log \left(\mu\right)-1\right)+O\left(\hbar ^2\right)\right)\\
+\left(
-\frac{\hbar }{12 \mu }+O\left(\hbar ^2\right) \right)
+\left( \frac{7 \hbar }{1440 \mu ^3}+O\left(\hbar ^2\right) \right)
+\left( -\frac{31 \hbar }{20160 \mu ^5}+O\left(\hbar ^2\right)  \right)
 +\cdots\,.
\end{multlined}
\end{equation}
We note that, in terms of $\mu$, all orders in $\hbar$ of $\widetilde \Pi_{DC}\left(E;\hbar\right)$ contribute at 1-loop order. 
The resummation of these terms leads to\footnote{This value can also be obtained by working directly with double turning points, as reviewed in appendix \ref{app:defquadratic}.}
\begin{equation}
\widetilde{a}_{DC}^{\text{\rm II}}(t;\hbar) = e^{-\frac{4}{3 \hbar }}
\left[\frac{\Gamma \left(t + 1\right)}{\sqrt{2\pi }}\right]^2 \left(\frac{8}{\hbar}\right)^{- 2\left(t+\frac{1}{2}\right) }\left(1+O(\hbar)\right).
\label{DC-Gamma-nu}
\end{equation}

Using \eqref{DC-Gamma-nu}, we can expand the ``symbol'' controlling the jump at $\arg\hbar = 0$ in \eqref{t-disc}. We find
\begin{equation}
\sqrt{\frac{\widetilde a_{DC}^{\text{\rm II}}}{\widetilde a_{AB}^{\text{\rm II}}}}\left(1+\widetilde a_{AB}^{\text{\rm II}}\right)
= e^{-\frac{2}{3 \hbar }}\frac{\sqrt{2 \pi }}{\Gamma (-t)}\left(\frac{8}{\hbar}\right)^{- \left(t+\frac{1}{2}\right) }\left(1+O(\hbar)\right).
 \label{eq:adcAsAS}
\end{equation}
Theorem 4.1.1 and Lemma 4.1.3 of \cite{dpham} imply that the resummation of the formal power series given by the $(1+O(\hbar))$ terms in \eqref{eq:adcAsAS} gives rise to a bounded function in $\hbar$ for sufficiently small $\hbar$, regular when $t$ is a positive integer. The results of \cite{dpham} apply close to a double turning point of general potentials, in particular generic polynomial potentials. In this way we have shown that the right hand side of \eqref{t-disc} vanish when the EQCs are imposed and \eqref{alien-swap} is verified. Borel summability of the energy eigenvalues is then proven for sufficiently small $\hbar$. Our proof has the same structure of the one in \cite{ddpham} but uses only single turning point techniques. Equation \eqref{eq:adcAsAS} corresponds to what is denoted $a^{[L]}$ in \cite{ddpham}.

\begin{figure}[t!]
\centering

\begin{subfigure}{0.3\textwidth}
\centering
\includegraphics[width=1\linewidth]{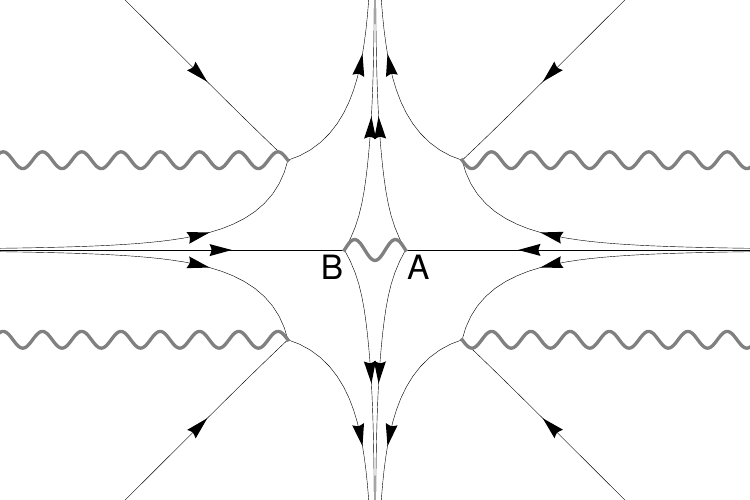}
\caption{$z^6+z^2$, $\arg[\hbar]=0$}
\end{subfigure}
\begin{subfigure}{0.3\textwidth}
\centering
\includegraphics[width=1\linewidth]{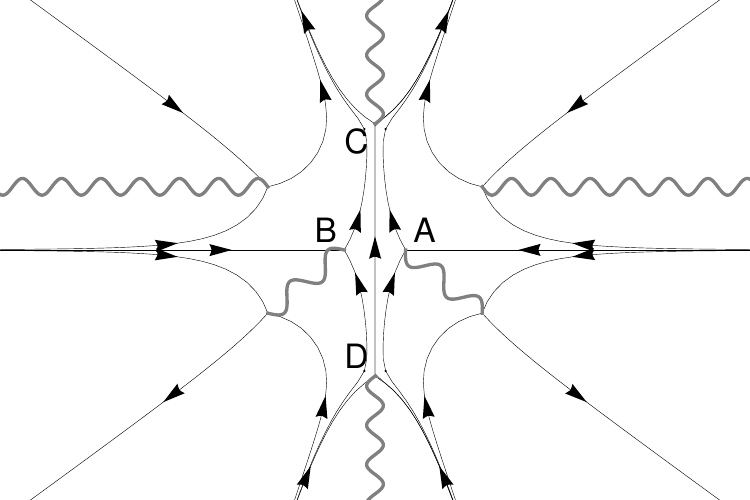}
\caption{$z^8+z^2$, $\arg[\hbar]=0$}
\end{subfigure}
\begin{subfigure}{0.3\textwidth}
\centering
\includegraphics[width=1\linewidth]{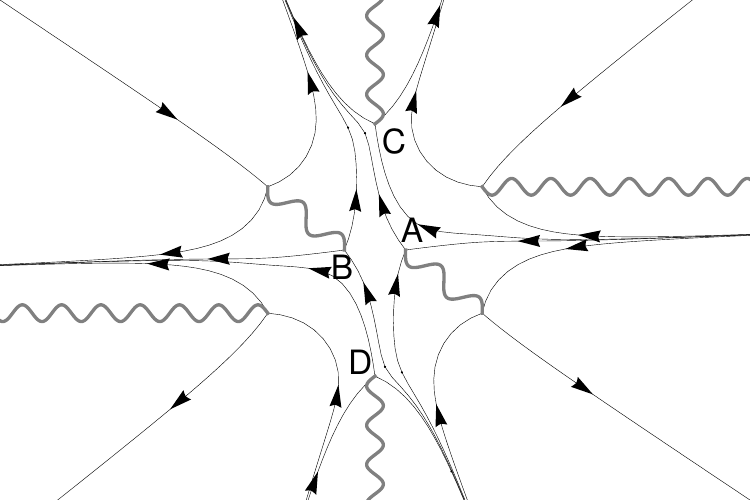}
\caption{$z^8+z^2$, $\arg[\hbar]=0.2$}
\end{subfigure}
\caption{Stokes graphs for higher order anharmonic potentials when $E$ is small.}
\label{fig:higher-order}
\end{figure}

\subsection{Higher order potentials}
\label{subsec:anharm-higher}

The considerations made for the quartic anharmonic potentials are easily generalizable to higher order potentials. As long as the potential is bounded with a single
global minimum $z_0$ there exist regions where the EQCs read as \eqref{simple-eqc}, with $a_{AB}$ the perturbative cycle. If we normalize the potential such that $V(z_0)=0$,
the ``sweet spot'' is found in the region where $\text{Arg} \,\hbar \gtrsim  c e_0$ for some model dependent constant $c$. For concreteness, 
we discuss the class of anharmonic potentials of the form
\be
V(z) =  \frac{1}{2}\left(z^2+z^{2q}\right)\,, \qquad   q>2\,.
\label{eq:PotAnhGen}
\end{equation}
The analysis is different for $q=even$ and $q=odd$.
For $q=odd$ the potentials \eqref{eq:PotAnhGen} cannot have purely imaginary turning points which lead to singular Stokes line crossing the perturbative cycle $a_{AB}$ at the origin $z_0=0$.
For sufficiently small $e_0$ and real $\hbar$, in fact no singular Stokes line crosses the cycle $a_{AB}$. Furthermore, there are no $\arg\hbar=0$ periods connecting to $a_{AB}$ and thus no Stokes jump. We are then automatically in region II where EQCs read as \eqref{simple-eqc} and
Borel summability of $a_{AB}$ is guaranteed. See figure \ref{fig:higher-order}, panel (a), for an illustration of the Stokes line configurations for $q=3$.

For $q=even$ the analysis is similar to the one of the quartic oscillator. For real $\hbar$, the non-perturbative cycle $a_{DC}$ turns into a singular Stokes line 
crossing $a_{AB}$. See figure \ref{fig:higher-order}, panel (b), for the $q=4$ case.
The complex $\hbar$ plane splits in several wedges with different EQCs. Crucially, we still have a region II like in the quartic case where the EQC reads \eqref{simple-eqc}.
This region is adjacent to region I as in figure \ref{fig-regs}, while the additional regions appear between regions II and III in figure \ref{fig-regs}.
This implies that the combined Stokes jumps between regions $\overline{{\rm II}}$ and II is as in section \ref{subsec:anharm-borel}, and \eqref{t-disc} applies.
The specific form of the series for the period $\widetilde\Pi_{DC}$ changes, but we still have
\begin{equation}
\sqrt{\frac{\widetilde a_{DC}^{\text{\rm II}}}{\widetilde a_{AB}^{\text{\rm II}}}}\left(1+\widetilde a_{AB}^{\text{\rm II}}\right)
= e^{-\frac{c_1}{\hbar }}\frac{\sqrt{2 \pi }}{\Gamma (-t)}\left(\frac{c_2}{\hbar}\right)^{- \left(t+\frac{1}{2}\right) }\left(1+O(\hbar)\right)\,,
 \label{eq:adcAsASG}
\end{equation}
where $c_1$ and $c_2$ are two model-dependent parameters. They are entirely determined from the leading and next to leading terms of the classical period. 
For the potentials \eqref{eq:PotAnhGen}, we have
\be
c_1=\frac{\sqrt{\pi}}{q+1} \frac{\Gamma \left(\frac{1}{q-1}\right)}{ \Gamma \left(\frac{1}{2}+\frac{1}{q-1}\right)} \,, \qquad c_2 = 2^{\frac{q+1}{q-1}}\,.
\label{eq:c1c2Def}
\ee
Note that \eqref{eq:adcAsASG} applies for any potential with a unique global minimum, not only for those of the kind \eqref{eq:PotAnhGen}. The universality of the factor
\be
\frac{\sqrt{2 \pi }}{\Gamma (-t)}\hbar^{t+\frac{1}{2} }
\label{eq:UniPidc}
\ee
is easily established in the strict double turning point limit $e_0=0$, where the factor \eqref{eq:UniPidc} appears in connection matrices of certain Stokes jump in double turning points \cite{dpham,ddpham}. See appendix \ref{app:defquadratic} for its derivation. The right hand side of \eqref{t-disc} then vanish when the EQCs are imposed and \eqref{alien-swap} is verified. The analysis can be generalized for an arbitrary polynomial potential with a unique global minimum.
Borel summability of the energy eigenvalues is then proven for sufficiently small $\hbar$.

\section{How EPT Borel summability emerges from EWKB}
\label{sec:HO}

In this section we show how to get Borel resummable series for energy eigenvalues of a general polynomial potential
even in the cases where we would expect a transseries expansion. We briefly review in section \ref{subsec:ept} the results of \cite{Serone:2016qog,power} and then show how the ``sweet spot'' regions for the anharmonic oscillators \eqref{eq:PotAnhGen} allows us to reproduce the findings of \cite{Serone:2016qog,power} in EWKB.
We finally discuss the leading large order behaviour of the EPT asymptotic series.

\subsection{Exact perturbation Theory (EPT)}
\label{subsec:ept}

It has been shown in \cite{Serone:2016qog,power} that for an arbitrary one-dimensional bounded polynomial potential $V$ we can define a perturbative
expansion such that all energy eigenvalues (and other observables as well) are reconstructable from a single Borel resummable perturbative series, dubbed EPT in \cite{Serone:2016qog,power}.
The idea is extremely simple and powerful at the same time. Let 
\be
V=V_0+\Delta V
\label{eq:Vdec}
\ee
be the sum of two potentials $V_0$ and $\Delta V$ such that $V_0$ has a single non-degenerate minimum and
\be
\lim_{|x|\rightarrow \infty} \frac{\Delta V}{V_0}= 0\,.
\label{eq:ept1C}
\ee
We take $\min(V) = \min(V_0) = 0$. Consider then the auxiliary potential 
\be
V_{\text{EPT}} = V_0+ \frac{\hbar}{\hbar_0}  \Delta V \equiv 
V_0+ \hbar V_1 \,, 
\label{eq:VEPTdec}
\ee
where $\hbar_0$ is an arbitrary positive constant. Using path integral methods and Lefschetz thimbles it has been shown in \cite{power} that the $\hbar$ expansion in the deformed quantum mechanical model with potential $V_{\text{EPT}}$, at fixed $\hbar_0$, is Borel resummable. 
Note that the condition \eqref{eq:ept1C} guarantees that the asymptotic behaviour of the wave function in the deformed model is the same as that in the original model. 
Let $\widetilde{E}^{\text{EPT}}_{n}(\hbar,\hbar_0)$ be the EPT asymptotic series
of the energy eigenvalues of the deformed model. Then the original energy eigenvalues $E_n$ can be computed as
\be
E_n(\hbar) = s_0\Big(\widetilde{E}^{\text{EPT}}_n(\hbar,\hbar_0)\Big)_{\hbar_0 = \hbar}\,.
\label{eq:EnEPT}
\ee
The decomposition \eqref{eq:Vdec} is generally far from being unique. EPT is defined as an expansion around the minimum of $V_0$,
which does not need to be a minimum of the original potential $V$. The number of interaction terms present in EPT depends on the particular decomposition performed. 
Any choice of EPT is theoretically equivalent to any other, though
in numerical computations with truncated series some choices might lead to more accurate results than others. 
EPT can also be used when the ordinary $\hbar$ expansion is Borel resummable, in order to improve the efficiency of the resummation at strong coupling.
Its more dramatic consequences however apply for quantum mechanical systems where, due to instanton configurations, energy eigenvalues are expected to be given
by the Borel resummation of a transseries in $\hbar$, $\exp(-1/\hbar)$ and possibly $\log \hbar$.

The numerical efficacy of EPT has been extensively discussed in \cite{Serone:2016qog,power} (see also \cite{Jaeckel:2018tdj}, and \cite{serone2} for an application in 2d QFT)
and will not be further considered in this paper, where we focus more on analytical aspects.
For this reason,  we will consider a specific decomposition \eqref{eq:VEPTdec} for concreteness. Given an arbitrary polynomial potential $V$ of the form
$V=\sum_{i=2}^{2q} v_i x^i$, normalized so that $v_{2q} = 1/2$, we take $V_0$ as in \eqref{eq:PotAnhGen}:
\be
V_0(z) =  \frac{1}{2}\left(z^2+z^{2q}\right)\,, \qquad q>1\,.
\label{eq:PotAnhGenEPT}
\end{equation}
In order to see how \eqref{eq:EnEPT} follows from an EWKB analysis requires a slight generalization of the
results of section \ref{sec:pure-anharmonic} to quantum deformed potentials $V_1$.
Note that all possible moduli of the original potential $V$ are encoded in $V_1$. The analysis in this section extends to non-polynomial $V_1$ as long it is polynomially bounded at infinity and regular on a infinitesimal neighbourhood of the interval $[-i,i]$. One such example is $z/\cosh(z)$. This naturally extends the applicability of EPT by broadening the class of original potentials $V$.

\subsection{Anharmonic oscillators in the presence of a quantum potential}
\label{sec:quant-anharm}

We assume that the quantum potential is a generic polynomial of degree strictly less than $2q$, so that condition \eqref{eq:ept1C} is satisfied.
Since the pattern of Stokes lines and jumps is determined from the classical potential, the analysis made in section \ref{sec:pure-anharmonic} applies.
In particular the EQC in region $\text{\rm II}$ is still given by \eqref{simple-eqc}. 
The discussion of Borel summability also carries over almost entirely from \ref{sec:pure-anharmonic}. 
The quantum periods are however affected by the quantum deformed potentials and hence we should verify that
\eqref{t-disc-noneven} vanishes when the EQC is imposed.

For even quantum potentials, at one-loop \eqref{t-disc-noneven} simplifies to \eqref{t-disc}, so we should determine the leading order form of $\widetilde a^{\rm II}_{DC}$ only. We get
\begin{equation}
\widetilde{a}_{DC}^{\text{\rm II}}(t;\hbar) = e^{-\frac{2c_1}{\hbar }}e^{2c_3}\left[\frac{\Gamma \left(t + 1\right)}{\sqrt{2\pi }}\right]^2 \left(\frac{c_2}{\hbar}\right)^{- 2\left(t+\frac{1}{2}\right) }\left(1+O(\hbar)\right)\,,
\label{DC-Gamma-nu2}
\end{equation}
where $c_1$ and $c_2$ are given by \eqref{eq:c1c2Def} and are unaffected by $V_1$, while the new real parameter $c_3$ in \eqref{DC-Gamma-nu2} reads  
\begin{equation}
c_{3} =\frac{1}{2}\lim_{e_0\rightarrow 0}\oint_{\gamma_{DC}} P_{{\rm even},1}(z) d z=i \int_{-1}^1 \frac{V_1(i t)}{\sqrt{Q_{0}(i t)}}\Bigg|_{E=0} d t\,,
\label{c3-quant}
\end{equation}
and is governed by $V_1$. For an arbitrary polynomial potential $V_1$, the integral in \eqref{c3-quant} is finite and $|c_3|$ is bounded.
Hence Borel summability is not affected.
For non-even quantum potentials, one obtains instead \eqref{t-disc-noneven}.
A calculation similar to the one leading to \eqref{DC-Gamma-nu} applies for the diagonal periods, such as
\begin{equation}
\widetilde{a}_{DA}^{\text{\rm II}}(t;\hbar) \approx e^{i\pi\left(t+\frac{1}{2}\right)} e^{-\frac{c_1}{\hbar }}e^{c_3-i\pi\vartheta}\frac{\Gamma \left(t + 1\right)}{\sqrt{2\pi }} \left(\frac{c_2}{\hbar}\right)^{- \left(t+\frac{1}{2}\right) }\left(1+O(\hbar)\right).
\label{Gamma-nu}
\end{equation}
Here $\vartheta$ stands for an additional contribution from non-even quantum potentials,
\begin{equation}
\vartheta= \frac{1}{2\pi i}\lim_{E\rightarrow 0}\oint_{\gamma_{AD}} \left\{P_{{\rm even},1}(z)-P_{{\rm even},1}(-z)\right\}d z = \frac{1}{\pi}\int_0^1 \frac{V_1(i t)-V_1(-i t)}{\sqrt{Q_{0}(i t)}}\Bigg|_{E=0}d t \,,
\end{equation}
where $P_{{\rm even},1}$ is the next to leading coefficient of $\widetilde P_{\text{even}}$ as defined in \eqref{eq:PevenoddV2}
More generally, we can use the symmetries of the problem to decompose the diagonal periods as
\begin{equation}
\begin{aligned}
\widetilde{a}_{CA} = \widetilde{a}_{DC}^{-\frac{1}{2}}\widetilde{a}_{AB}^{-\frac{1}{2}}
e^{-i\pi\widetilde\vartheta(\hbar)},\quad 
\widetilde{a}_{BD} = \widetilde{a}_{DC}^{-\frac{1}{2}}\widetilde{a}_{AB}^{-\frac{1}{2}}
e^{i\pi\widetilde\vartheta(\hbar)},
\\
\widetilde{a}_{DA} = \widetilde{a}_{DC}^{\frac{1}{2}}\widetilde{a}_{AB}^{-\frac{1}{2}}
e^{-i\pi\widetilde\vartheta(\hbar)},\quad
\widetilde{a}_{BC} = \widetilde{a}_{DC}^{\frac{1}{2}}\widetilde{a}_{AB}^{-\frac{1}{2}}
e^{i\pi\widetilde\vartheta(\hbar)},
\end{aligned}
\label{eq:all-diagonal}
\end{equation}
where
\begin{equation}
\widetilde\vartheta(\hbar)=\frac{1}{2\pi i\hbar}\lim_{E\rightarrow 0}\int_{AD}\left\{\widetilde P_{\rm even}(z)-\widetilde P_{\rm even}(-z)\right\}dz \sim \vartheta + \CO(\hbar).
\end{equation}
It can be shown that $\widetilde\vartheta(\hbar)$ is real to all orders.
Putting everything together, \eqref{t-disc-noneven} gives rise to
\begin{equation}
s_0^{-1} \lim_{e_0\rightarrow 0} \Big(t^{\rm II}-t^{\overline{\rm II}}\Big) = 
\frac{1}{2\pi i}
\bigg\{
\log\Big(1+e^{-\frac{c_1}{\hbar }}e^{c_3-i\pi\vartheta}\frac{\sqrt{2 \pi }}{\Gamma (-t)}\left(\frac{c_2}{\hbar}\right)^{- \left(t+\frac{1}{2}\right) }\left(1+O(\hbar)\right)\Big)+
{\rm h.c.}
\bigg\} \,,
\label{borel-ept}
\end{equation}
and vanishes when EQC is imposed, $t=n$. It then follows from \eqref{borel-ept} that 
\begin{equation}
\dot{\Delta}_A \widetilde E^{\text{EPT}}_n = 0, \quad \forall A \in \mathbb{R}^+ \,,
\label{alien-swap-EPT}
\end{equation}
and hence Borel summability of the EPT series of energy eigenvalues for sufficiently small $\hbar$ is established.

\subsection{Large order behaviour for quartic oscillators with quantum potentials}
\label{subsec:LOBq}

We discuss in this section the large order behavior of the coefficient terms of the  EPT asymptotic series $\widetilde E^{\text{EPT}}_n=\sum_k e^{\text{EPT}}_{k,n}\hbar^k$. For simplicity we focus on quartic oscillators.  In absence of a quantum potential $V_1$, the large order behaviour associated to the potential \eqref{eq:PotAnh} has been famously found long ago by Bender-Wu \cite{Bender:1973rz}
and more rigorously established using alien calculus in \cite{ddpham}. As well-known, all Borel singularities are along the negative real axis.  
We extend here the analysis of \cite{ddpham} to include a broad class of quantum potentials, so that the large order behaviour of $\widetilde E^{\text{EPT}}_n$ follows. 

Like in the analysis of Borel summability, we need to study in more detail what happens to the non-perturbative periods after applying the EQCs.
To obtain the large order behavior in the $e_0\rightarrow 0$ limit we must compare regions $\text{III}$ and $\overline{\text{III}}$, which in the vanishing limit correspond respectively to $s_{\pi-}$ and $s_{\pi+}$. Using \eqref{eq:DDP}, we derive
\begin{equation}
a_{AB}^{\overline{\text{III}}}=\left(1+(1+a_{AB}^{\text{III}})a_{CA}^{\text{III}}\right)\left(1+(1+a_{AB}^{\text{III}})a_{BD}^{\text{III}}\right)a_{AB}^{\text{III}}.
\end{equation}
Unlike the jump for $\arg\hbar=0$ studied in the previous sections where all terms were regular, here $a_{CA}^{\text{III}}$ is in fact singular when we replace $t$ by $n$ (equivalently, $E_1$ by $\widetilde{E}_n$).
To see why, consider the jump at $\arg\hbar=\pi/2$,
\begin{equation}
a_{CA}^{\text{III}}=\left(1+a_{AB}^{\text{\rm II}}\right)^{-1}a_{CA}^{\text{\rm II}}.
\label{eq:i-jump-E}
\end{equation}
While it is a non-perturbative term when expressed in terms of $(E;\hbar)$, it becomes an overall coefficient for series parametrized by $t$. After changing variables in \eqref{eq:i-jump-E} from $E$ to $t=t^{\rm II}$ and then expanding at small $\hbar$ (i.e. ``downgrading'' it), we find
\begin{equation}
\widetilde{a}_{CA}^{\text{III}}(E(t;\hbar);\hbar)=\left(1-e^{-2\pi i t}\right)^{-1}\widetilde{a}_{CA}^{\text{\rm II}}(E(t;\hbar);\hbar).
\end{equation}
Notice that, due to the change variables from $E$ to $t$, ``perturbative'' and ``non-perturbative'' are rearranged and the asymptotic series in the different regions differ.
In the $e_0\rightarrow 0$ limit we obtain, to leading non-perturbative order,
\be
\begin{split}
s_\pi^{-1} \lim_{e_0\rightarrow 0} \Big(t^{\overline{\rm III}}- t^{\rm III}\Big) & = \frac{1}{2\pi i} \log\Big(\left(1+\tilde{a}_{CA}^{\text{\rm II}}\right)\left(1+\tilde{a}_{BD}^{\text{\rm II}}\right)\Big)   \\
& =  \cos (\pi  \vartheta )  e^{\frac{2}{3 \hbar }}\frac{\sqrt{2\pi }}{ \pi i\Gamma \left(t +1\right)}\left(\frac{8 e^{i\pi}}{\hbar}\right)^{ \left(t+\frac{1}{2}\right) }e^{-c_3}\left(1+O(\hbar)\right) +\cdots \,, 
\label{stokes-pi}
\end{split}
\ee
from which we can extract the large order behavior:\footnote{This is obtained by expanding both the Stokes automorphism in the left hand side of \eqref{stokes-pi} in terms of dotted alien derivatives using \eqref{app:StokesDef2} and the log in the right hand side of \eqref{stokes-pi}, matching terms order by order in $\exp(2p/(3\hbar))$, for $p\in \mathbb{N}$. We determine in this way $\dot{\Delta}_A \widetilde E^{\text{EPT}}_n$ for $A = -2p/3 \in \mathbb{R}^-$ and then the large order behaviour \eqref{large-order} (as well as higher order terms) follows from standard resurgence relations (see e.g.  \cite{Dorigoni:2014hea,abs}).}
\begin{equation}
e^{\text{EPT}}_{k,n}\sim (-1)^{k+1} e^{-c_3}\cos (\pi  \vartheta ) \frac{   2^{3 n+1}}{  \pi^{3/2} }\frac{\Gamma \left(k+n+\frac{1}{2}\right)}{\Gamma (n+1)}\left(\frac{3}{2}\right)^{k+n+\frac{1}{2}}.
\label{large-order}
\end{equation}
In the absence of a quantum potential, formula \eqref{stokes-pi} matches the equivalent formula in  \cite{ddpham},
where it corresponds to $a^{\cal L}$ in the notation therein, and \eqref{large-order} becomes the Bender-Wu formula.

The quantum potential affects the leading large order behaviour only in modifying the overall constant factor through the coefficients $c_3$ and $\theta$.
Non-perturbative effects in EPT are hidden in the negative real axis of the Borel plane and
are essentially encoded in $c_3$ and $\theta$, as well as higher order corrections, which in general depend on $\hbar_0$.
An illustrative example of the above formula is the EPT model:
\begin{equation}
V_0(z) = \frac{1}{2}\left(z^2+z^4\right),\quad V_{1}(z) = a z - \frac{z^2}{\hbar_0}\,.
\label{eq:ept-eg}
\end{equation}
This model has the anharmonic oscillator as its classical ``base'', but can be used to reproduce some interesting test cases. For $a=0$, at $\hbar_0=\hbar$ it reduces to the symmetric double well, while at $\hbar_0=2\hbar$ it becomes the pure quartic potential. Meanwhile if $a=1/2$ and $\hbar_0=\hbar$ it becomes the \textit{supersymmetric} double well.
The perturbative series for the energy levels is given by
\begin{align}
\widetilde{E}^{\text{EPT}}_n & =\left(n+\frac{1}{2}\right)\hbar + \left( -\frac{a^2}{2}-\frac{ \left(n+\frac{1}{2}\right)}{\hbar _0}+\frac{3}{8} (2 n (n+1)+1)\right)\hbar^2 \\
& -\left((2 n+1)  \left(-48 a^2+17 n (n+1)+21\right)+\frac{8}{\hbar_0} \left(4 a^2-6 n (n+1)-3\right)+\frac{16 n+8}{32 \hbar _0^2}\right)\hbar^3+\cdots  \,. \nn
\end{align}
In the EPT model \eqref{eq:ept-eg} the coefficients $c_3$ and $\theta$ read
\begin{equation}
e^{-c_3} = e^{-\frac{2}{\hbar_0}},\quad \vartheta = a.
\label{large-order-eg}
\end{equation}
We explicitly see that when we set $\hbar_0 =\hbar$, the leading singularity at $2/3$ becomes multiplied by the non-perturbative factor $e^{-2/\hbar}$
in the double-well potential and the quartic oscillator.
As for the supersymmetric potential, we see that even at the level of EPT the value of $a=1/2$ is special, since it puts the leading singularity further away. The next contribution to the large behavior is 
\begin{equation}
e^{\text{EPT}}_{k,n}\approx (-1)^{k+1} e^{-\frac{4}{h_0}} \frac{ 4^{3 n+1}  }{\pi  }\frac{  \Gamma (k+2 n+1)}{  \Gamma (n+1)^2}\left(\frac{3}{4}\right)^{k+2 n+1},
\end{equation}
which follows from \eqref{stokes-pi}. Consequently, the EPT perturbative series never vanishes, in contrast with the ordinary perturbative series in supersymmetric quantum mechanics. This is expected, since the EPT perturbative series must capture the non-perturbative effects that give rise to $E_{n=0}> 0$ in the exact case.\footnote{$E_{n=0}$ is the exact ground state energy and it should not be confused with $e_0$ defined in \eqref{eq:EtildeAN}.}

\section{Conclusions}
\label{sec:conclu}

We have shown in this paper how the results of \cite{Serone:2016qog,power} about EPT, obtained using path integral and steepest-descent  considerations,
nicely fit with resurgence and EWKB methods, providing an alternative (and more rigorous) proof of the validity of EPT which applies to all energy eigenvalues at once. 
EPT in \cite{Serone:2016qog,power} is based on the existence of deformations for which the Lefschetz thimble decomposition of the path integral trivializes and correspondingly
observables can be expressed in terms of a single Borel resummable asymptotic series.
In EWKB, at fixed moduli of the potential, EQCs depend in particular on $E$ and $\arg \hbar$. In theories where instanton contributions occur, there is no way to get
a EQC of the form \eqref{eq:EQCsimple}. Non-perturbative cycles necessarily enter the EQCs and in particular take into account of the instanton contributions.
The expansion of an observable in $\hbar$ in this case is generally given by a transseries.
However, the very same deformations above, provided $E$ is carefully chosen, are able to lead to a trivialization of the EQC and to a single Borel resummable asymptotic series
for energy eigenvalues. 

We have focused in this paper to polynomial potentials and it would be interesting to extend our results to more general potentials. 
It would also be interesting to see if and to what extent one can reconstruct the transseries expression for an observable (in terms of the undeformed model)
by unzipping its associated EPT series.
As we mentioned in the introduction, connections between EWKB and several other theoretical subjects have been worked out. It would be very interesting to
see if EPT in EWKB can be extended in this more general context, in particular in the context of 4d ${\cal N}=2$ gauge theories.

\section{Aknowledgments}

We thank Alba Grassi, Alexander van Spaendonck and Andr\'e Voros for discussions. 
We thank the organizers of the 2022 workshop ``Physical resurgence: On quantum, gauge, and stringy" held 
at the Isaac Newton Institute in Cambridge and the 2023 workshop ``Quantization and Resurgence" held at the SwissMAP Research Station in Les Diablerets for the hospitality, and the participants of both workshops for interesting discussions.
MS thanks Veronica Fantini and Maxim Kontsevich for discussions, Yoshitsugu Takei for a useful e-mail correspondence, Syo Kamata for clarifications on some of the results of \cite{Sueishi:2021xti}. BB thanks SISSA for the hospitality while carrying out part of this work.
MS thanks the Institut des Hautes \'Etudes Scientifiques (IHES) for the hospitality during the completion of this project.  
TR is partially supported by the ERC-COG grant NP-QFT No.~864583 ``Non-perturbative dynamics of quantum fields: from new deconfined phases of matter to quantum black holes'', by the MIUR-SIR grant RBSI1471GJ, by the MIUR-PRIN contract 2015 MP2CX4.
BB is partially supported by INFN Iniziativa Specifica TPPC. TR and MS are partially supported by INFN Iniziativa Specifica ST\&FI.
\appendix

\section{Connection matrices for monomial potentials}
\label{app:connmat}

We compute in this appendix the connection matrices for wave functions as they cross a regular Stokes line. For simple turning points
this connection matrix is given by the Airy function case with $Q(z)=z$.  
Since it is straightforward to generalize to pure monomial potentials, 
we report the connection matrices for potentials $Q(z)=z^q$, with $q\in \mathbb{N}$.  It should be stressed that such general connection matrices with $q>1$
are not directly useful to determine the connection matrices for higher order turning points in a {\it generic} potential. Determining such matrices is in fact a non-trivial task.

The differential equation reads
\be
\hbar^2 \psi^{\prime\prime}(z)  =  z^q \psi(z)\,, \qquad q\in \mathbb{N}\,.
\label{ODGp1}
\ee
The leading terms of the expansion are given by a WKB approximation:
\be
\widetilde \psi_\pm \approx z^{-\frac q4} e^{\pm \frac{z^{\alpha}}{\alpha\hbar} }\,, 
\label{ODGp3}
\ee
where 
\be
\alpha = \frac{q+2}{2}\,.
\ee
We look for asymptotic solutions of the form
\be
\widetilde \psi_\pm = z^{-\frac q4} e^{\pm \frac{z^{\alpha}}{\alpha\hbar} }  \widetilde\phi_\pm\,, \qquad  \widetilde\phi_\pm = \sum_{n=0}^\infty c_n^{\pm}(z) \hbar^n \,,
\label{ODGp4}
\ee
where $c_0(z)=1$. We have 
\be
c_n^\pm(z) = d_n^\pm \,  z^{-n \alpha}\,,
\label{ODGp5}
\ee
and after some algebra we get the recursion relation satisfied by $d_n^\pm$:
\be
d_{n+1}^\pm = \frac{\pm d_n^\pm}{(n+1)(q+2)} \bigg[\frac q4 \Big(\frac q 4+1\Big) + n(n+1)  \Big(\frac q 2+1\Big)^2\bigg] \,,
\label{ODGp5a}
\ee
whose solution is 
\be
d_n^{\pm} = \sin\Big(\frac{\pi q}{2q+4}\Big)  \Big(\pm \frac{q+2}{4}\Big)^n \frac{ \Gamma\Big(n+\frac{q}{2(q+2)}\Big)\Gamma\Big(n+\frac{q+4}{2(q+2)}\Big)}{\pi \Gamma(n+1)}\,.
\label{ODGp6}
\ee
The Borel transforms of $\widetilde \phi_\pm$ read
\be
\widehat \phi_{\pm}(\tau) =   {}_2F_1\Big(\frac{q}{4+2q}, \frac{4+q}{4+2q}, 1; \pm \frac{q+2}{4} \tau \Big)\,, \qquad \tau = t \,z^{-\alpha}\,.
\label{ODGp7}
\ee
The actual expansion parameter of the formal series in \eqref{ODGp4} is $\hbar z^{-\alpha}$ and correspondingly $\tau$ in \eqref{ODGp7} is the associated Borel variable. 
The singularities occur at $\tau_+=4/(q+2)$ for $\widehat \phi_{+}(\tau)$ 
and $\tau_- =-4/(q+2)$ for $\widehat \phi_{-}(\tau)$. 
If $z = |z| \exp(i \theta)$, with $\hbar$ and $t$ real and positive, Stokes lines occur for 
\be
\theta = \frac{2\pi}{q+2} n\,, \qquad n=0,1,\ldots, q+1\,,
\label{ODGp8}
\ee
and the complex $z$-plane splits in $q+2$ wedges with opening angle $2\pi/(q+2)$. The Borel resummation in a generic wedge X is defined as 
\be
\phi_\pm^\text{X}  = s_\theta(\widehat \phi_\pm)\,, \qquad z\in \text{X}\,.
\ee
The expansion of $\widehat\phi_\pm$ around $\tau_\pm $ equals
\begin{align}
\widehat \phi_\pm |_{\tau = \tau_\pm} & =- \frac{\eta_q}{2\pi} \log (\tau -\tau_\pm)\widehat \phi_\mp |_{\tau = 0} + \text{reg}\,, \qquad \eta_q =  2\cos\Big(\frac{\pi}{q+2}\Big)\,,
\label{ODGp9}
\end{align}
where reg stands for analytic terms in a neighborhood of $\tau_\pm$. Let $\phi^{\rm I}_\pm$ and $\phi^{\rm II}_\pm$  be the Laplace transforms of $\widehat \phi_\pm$ in two adjacent wedges separated by a Stokes line for $\widehat \phi_+$.
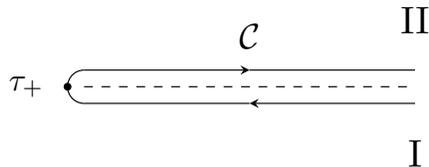
\begin{figure}[t!]
    \centering
    \raisebox{-3.em}
 {\scalebox{1.1}{
    \begin{tikzpicture}

\filldraw[fill=white, draw=black] (7,1/5) arc (90:270:1/5);
\draw (7,-1/5) - - (9,-1/5);
\draw   [stealth-]  (9,-1/5) - - (11,-1/5);
\draw  [-stealth] (7,1/5) - -  (9,1/5);
\draw (9,1/5) - -  (11,1/5);
\draw   [dashed]  (7,0) - - (11,0);
         \filldraw[fill=black, draw=black]  (6.8,0) circle (0.04 cm);

  \node at (6.3,0) {\scalebox{1.}{$\tau_+$}};
      \node at (9.,0.6) {\scalebox{1.1}{${\cal C}$}};
    \node at (11.,0.8) {\scalebox{1.2}{\rm II}};
        \node at (11.,-0.8) {\scalebox{1.2}{\rm I}};
\end{tikzpicture}}}
    \caption{Hankel contour passing through the singular point $\tau_+$. The dashed line delimits the two adjacent regions II and I in the complex  $z$-plane.}
   \label{fig:Hankel}
\end{figure} 
We have
\begin{align}
\phi_+^\text{\rm II}(z) - \phi_+^\text{\rm I}(z) & =  \Big(-\frac{\eta_q}{2\pi \hbar}\Big) \int_{{\cal C}}\! e^{-\frac t \hbar} \log\Big(\tau -\tau_+\Big) \widehat\phi_-(\tau-\tau_+) \\
& = \frac{i \eta_q}{\hbar} \int_{\tau_+}^\infty \!\! dt\, e^{-\frac{t}{\hbar}} \widehat \phi_-(\tau-\tau_+) = e^{-\frac{2z^\alpha}{\alpha\hbar }} \frac{i \eta_q}{\hbar} \int_{0}^\infty \!\! dt\, e^{-\frac{t}{\hbar}} \widehat \phi_-(\tau)\,, \nn
\end{align}
where ${\cal C}$ is the Hankel contour shown in figure \ref{fig:Hankel}.
Hence
\be
\psi_+^\text{\rm II} - \psi_+^\text{\rm I} =  i\eta_q   z^{-\frac q4} e^{-\frac{z^\alpha}{\alpha\hbar}}\frac{1}{\hbar} \int_0^\infty \! dt \, e^{-\frac{t}{\hbar}} \widehat \phi_-(\tau) =
i \eta_q \psi_-^\text{\rm I}\,.
\label{ODI13}
\ee
Since $\psi_-^\text{\rm I}$ is regular when $\psi_+$ jumps, we have
\be
\psi_-^\text{\rm II} = \psi_-^\text{\rm I} \,.
\label{ODI13a}
\ee
A similar analysis applies in two adjacent wedges separated by a Stokes line for $\widehat \phi_-$. In that case we get
\be
\begin{split}
\psi_-^\text{\rm II} - \psi_-^\text{\rm I}  & = i \eta_q \psi_+^\text{\rm I} \,, \\
\psi_+^\text{\rm II} & = \psi_+^\text{\rm I} \,.
\label{ODI13b}
\end{split}
\ee
We attach an arrow to Stokes lines to distinguish the two cases. An arrow entering (exiting) the turning point corresponds to Stokes lines where $\psi_+$ ($\psi_-$)  jumps.
We have also to take into account from \eqref{ODGp4} the presence of the branch-cut singularity at $z=0$, which is different depending on whether $q$ is even or odd.
In terms of the coefficients defined in \eqref{eq:cDef}, we get 
\be
B^{(q)}= e^{- \frac{i \pi q}{2}}\frac 12 \left(\begin{array}{cc}
1+(-1)^q  & 1-(-1)^q   \\
1-(-1)^q  & 1+(-1)^q 
\end{array}\right) \,,
\label{ODGp14b}
\ee
whereas the Stokes jumps \eqref{ODI13}-\eqref{ODI13b} are encoded in the connection matrices 
\be
S_+^{(q)} = \left(\begin{array}{cc}
1 & 2i\cos\Big(\frac{\pi}{q+2}\Big) \\
0 & 1
\end{array}\right) \,, \qquad 
S_-^{(q)} = \left(\begin{array}{cc}
1 & 0  \\
2i\cos\Big(\frac{\pi}{q+2}\Big) & 1
\end{array}\right) \,.
\label{ODGp14a}
\ee
For integer $q$, solutions of the differential equations are analytic over the complex plane and so under a full rotation over the complex $z$-plane 
\be
\psi_\pm^\text{X}(e^{2i \pi }z) = \psi_\pm^\text{X}(z) \,, 
\ee
where X is the wedge of the complex plane which includes the starting point $z$. Under a $2\pi$ rotation, the total monodromy is given by the product of the $q+2$ connection matrices
\eqref{ODGp14a} and the branch-cut  matrix in \eqref{ODGp14b}. In total we indeed get
\be
(S_+^{(q)} S_-^{(q)})^{[\frac{q+2}{2}]} (S_+^{(q)})^{\frac{1-(-1)^q}{2}}  B^{(q)} = I\,, \qquad \forall q\in \mathbb{N}\,,
\label{eq:ODGp15}
\ee
where $[x]$ is the integer part of $x$. The result does not depend on the location of the branch cut.
For even $q$, $B^{(q)} = (-1)^{q/2}$, while 
\be
S_\pm^{(q)} B^{(q)} = B^{(q)} S_\mp^{(q)}\,, \qquad \text{odd} \; q\,.
\label{eq:Bscomm}
\ee
The most relevant case is $q=1$, so for simplicity we define
\be
S_+ \equiv S_+^{(1)}\,, \qquad  S_- \equiv S_-^{(1)}\,, \qquad B\equiv B^{(1)}\,,
\ee
which are the matrices \eqref{eq:monodromy} in the main text.

Define the Stokes automorphism 
\be
\mathfrak{S}_\theta \equiv s_{\theta^-}^{-1} \cdot s_{\theta^+}\,,
\label{app:StokesDef1}
\ee
where $\theta^\pm = \theta\pm \epsilon$, $\epsilon\ll1$, and $s_{\theta^\pm}$ represent the lateral Borel resummations defined in \eqref{eq:BorelDef}. 
In our case the two non-trivial points where a Stokes jump occurs are $\tau=\tau_+$ and $\tau=\tau_-$, corresponding respectively to  $\theta_{\tau_+} =2\pi (2n)/(q+2)$
and $\theta_{\tau_-} =2\pi (2n+1)/(q+2)$ in the $z$-plane, with $n$ an integer. 
In terms of $\widetilde\psi_\pm$, we get
\be
\begin{split}
\mathfrak{S}_{\theta_{\tau_+}} \widetilde \psi_+ & = \widetilde \psi_+  + i \eta_q \,\widetilde \psi_- \,, \\
\mathfrak{S}_{\theta_{\tau_-}} \widetilde \psi_- & = \widetilde \psi_- +  i \eta_q \,\widetilde  \psi_+\,,
\label{eq:StokesAiry}
\end{split}
\ee
so that the connection formulas for the wave functions can be interpreted in terms of Stokes automorphisms of the associated formal power series.
The Stokes automorphism can also be defined in terms of alien derivatives (dotted and undotted versions):
\be
\mathfrak{S}_\theta \equiv \text{exp}\Big(\sum_{\{\omega_\theta\}} \dot \Delta_{\omega_\theta}  \Big)\,,
\qquad \dot \Delta_{\omega_\theta} \equiv e^{-\frac{\omega_\theta}{\hbar}} \Delta_{\omega_\theta}\,,
\label{app:StokesDef2}
\ee
where $\{\omega_\theta\}$ are all the possible singularities (just one in our case) along the ray with angle $\theta$.
If $\omega$ is an analytic point of the Borel transform 
associated to an asymptotic series function $\widetilde f(\hbar)$, then
$\Delta_\omega \widetilde f(\hbar) = 0$. It easily follows from \eqref{eq:StokesAiry} that 
 \begin{align}
\Delta_{\tau_+} \widetilde \phi_+ & = i  \eta_q\, \widetilde \phi_- \,, \qquad \,\Delta_{\tau_+} \widetilde  \phi_- = 0 \,, \nn \\
\Delta_{\tau_-} \widetilde  \phi_+ & = 0\,,  \qquad \quad\quad \; \Delta_{\tau_-}  \widetilde \phi_- = i \eta_q\, \widetilde  \phi_+\,.
\end{align}

\section{Double turning points}
\label{app:defquadratic}

In contrast to simple zeros, whose connection matrices can be reduced to that of the Airy function, double zeroes are more complicated. Intuitively it is simple to understand the origin of the complication. A double turning point can be seen as the annihilation of two simple turning points as we move parameters in the function $Q(z)$. No matter how close the turning points are, the Voros symbol $a_{\gamma_{\text{vc}}}$ associated to the vanishing cycle $\gamma_{\text{vc}}$
will in general be non-vanishing. In contrast to simple turning points, we then expect that the connection matrices of double turning points cannot be universal but depend
on a variable $\lambda$ defined as 
\be
a_{\gamma_{\text{vc}}}  \equiv  e^{-2i \pi (\lambda+\frac 12)}\,.
\label{eq:VorosTmono}
\ee
It has been shown in \cite{dpham,ddpham} (see also \cite{AKT1,AKT2}) that the full connection matrices for double turning points split into
a universal and a model-dependent term. The universal term coincides with the connection matrices for the Weber model defined by the equation
\be
\hbar^2 \psi^{\prime\prime}(z)  = \Big( \frac{z^2}{4} - \hbar \Big(\lambda+\frac 12 \Big) \Big) \psi(z)\,.
\label{ODQG1}
\ee
Aim of this appendix is to derive the connection matrices for the Weber model \eqref{ODQG1}. Although this is a known result, we believe it can be useful
to report here a comprehensive and detailed derivation aimed at physicists.\footnote{A similar, but less detailed, derivation appears in \cite{Sueishi:2021xti}. In particular 
the constraints given by imposing trivial total monodromy of the wave function are not considered there.}

The leading terms of the WKB expansion are found from \eqref{eq:Pcoeff}. We have
\be
P_0^\eta = \eta \frac{z}{2} \,, \qquad P_1^\eta = \frac{1}{2z}-\eta \frac{1+2\lambda}{2z}\,,
\label{ODQG2}
\ee
and hence 
\be
\widetilde \psi_\pm = z^{-\frac 12\mp (\lambda+\frac 12)} e^{\pm \frac{z^2}{4\hbar}}  \widetilde \phi_\pm\,, \qquad  \widetilde \phi_\pm = \sum_{n=0}^\infty  c_n^{\pm}(z) \hbar^n\,,
\label{ODQG3}
\ee
where $c_0(z)=1$. It is easy to see that for any integer $n$
\be
c_n^\pm(z) = d_n^\pm \,  z^{-2n}\,.
\label{ODQG4}
\ee
Plugging the ansatz \eqref{ODQG3} in \eqref{ODQG1} allows us to fix all the coefficients $d_n^\pm$. We get the following recursion relations for $d_n^\pm$:
\be
d_{n+1}^+ =  \frac{2d_n^+}{n+1} \Big(n+\frac{\lambda+1}2\Big) \Big(n + \frac{\lambda+2}{2}\Big)\,, \qquad
d_{n+1}^- = - \frac{2d_n^-}{n+1} \Big(n-\frac{\lambda}2\Big) \Big(n - \frac{\lambda-1}{2}\Big)\,, 
\ee
which are solved for
\be
d_n^{+} =  \frac{2^{n+\lambda}}{\sqrt{\pi}} \frac{\Gamma\Big(n+\frac{\lambda+1}{2}\Big) \Gamma\Big(n+\frac{\lambda+2}{2}\Big)}{\Gamma(n+1)\Gamma(\lambda+1)}  \,,\quad 
d_n^{-}  = \frac{2^{n-\lambda-1}}{\sqrt{\pi}} (-1)^n \frac{\Gamma\Big(n-\frac \lambda 2\Big) \Gamma\Big(n-\frac{\lambda-1}{2}\Big)}{\Gamma(n+1)\Gamma(-\lambda)} \,.
\label{ODQG5}
\ee
We compute the Borel transforms of $\widetilde \phi_\pm$. Since the powers of $z$ in front of $\widetilde \psi_+$ and $\widetilde\psi_-$ in the first relation in \eqref{ODQG3} are different, it is useful to define more general Borel-Leroy transforms:
\begin{align}
\hat\phi_{+}^{a_+}(\tau) & =  \sum_{n=0}^\infty \frac{c_n^+(z) }{\Gamma(n+1+a_+)} t^n = \frac{1}{\Gamma(1+a_+)}\,
{}_2F_1\Big(\frac{1+\lambda}2,\frac{2+\lambda}{2},1+a_+;\tau\Big) \,,  \nn \\
\hat\phi_{-}^{a_-}(\tau) & =  \sum_{n=0}^\infty \frac{c_n^-(z) }{\Gamma(n+1+a_-)} t^n = \frac{1}{\Gamma(1+a_-)}\,
{}_2F_1\Big(\frac{1-\lambda}2,-\frac{\lambda}{2},1+a_-; - \tau\Big) \,, 
\label{ODQG6}
\end{align}
where 
\be
\tau =\frac{2t}{z^2}\,,
\ee
and $a_\pm$ are parameters which we conveniently fix as explained below.
The only singularities occur at $\tau=1$ for $\hat\phi_{+}(\tau)$ and $\tau =-1$ for $\hat\phi_{-}(\tau)$. 
When $a\neq 0$, \eqref{eq:BorelDef} generalizes to 
\be
f^\theta(\hbar)  = \frac{1}{\hbar}  \int_0^{e^{i \theta \infty}} \!\! \! dt \, \hat f^a(t)  \Big(\frac{t}{\hbar}\Big)^a\, e^{-\frac t\hbar} \,.
\ee
If $z = |z| \exp(i \theta)$, with $\hbar$ and $t$ real and positive, Stokes lines occur for 
\be
\theta = 0, \frac{\pi}{2},\pi,\frac{3\pi}{2}\,.
\ee
The computation of the discontinuity $s_{\theta^+} - s_{\theta^-}$ is more involved than that for pure monomials.
In particular, care should be paid in keeping track of phases. 
We first consider $(s_{0^+} - s_{0^-})\psi_+$,  with $z$ taken real and positive so that no extra phases appear. In other words, the discontinuity is expected to be purely imaginary (like in the Airy case)
for any real value of $\lambda$. 
First of all we determine a convenient choice of $a_\pm$ which simplifies the computation. This is achieved by 
the use of the hypergeometric identity
\begin{equation}
\begin{aligned}
{ }_2 F_1(a, b , c, z)=& \frac{\Gamma(c) \Gamma(c-a-b)}{\Gamma(c-a) \Gamma(c-b)}{ }_2 F_1(a, b , a+b+1-c , 1-z) \\
&+\frac{\Gamma(c) \Gamma(a+b-c)}{\Gamma(a) \Gamma(b)}(1-z)^{c-a-b}{ }_2 F_1(c-a, c-b , 1+c-a-b , 1-z)\,,
\label{eq:hypidentity}
\end{aligned}
\end{equation}
which immediately shows that the non-analytic term of the expansion around $z=1$ is related to the expansion around $z=0$ of another hypergeometric function.
Demanding that the non-analytic behaviour of $\hat\phi_+^{a_+}$ around $\tau=1$ is governed by $\hat\phi_-^{a_-}$ around $\tau=0$ allows us to uniquely fix the coefficients $a_\pm$ by means of  \eqref{eq:hypidentity}. We get $a_+ = 0$, $a_- =- \lambda - 1/2$, resulting in 
\be
\hat\phi_+(\tau)\Big|_{\tau=1} =\kappa(\lambda) (1-\tau)^{-\lambda-\frac 1 2} \hat\phi_-(\tau-1) + \text{regular}\,,
\label{eq:ODQG6b}
\ee
where $\hat\phi_+ \equiv \hat\phi_+^{0}$,  $\hat\phi_- \equiv \hat\phi_-^{-\lambda-1/2}$, and
\be
 \kappa(\lambda) =  \frac{\Gamma\Big(-\lambda-\frac 12\Big)\Gamma\Big(-\lambda+\frac 12\Big)}{\Gamma\Big(\frac{1+\lambda}2\Big) \Gamma\Big(\frac{2+ \lambda}2\Big)}\,.
\ee
Taking
\be
1-\tau = e^{-i \pi} (\tau-1)\,, \qquad z = \rho \in \mathbb{R}
\ee
we have
\begin{align}
(s_{0^+} -s_{0^-})\widetilde \psi_+  & = \rho^{-\lambda-1} e^{\frac{\rho^2}{4\hbar}}e^{i \pi (\lambda+\frac 12)} \Big(1- e^{-2i \pi (\lambda+\frac 12)} \Big) \kappa(\lambda)
\frac 1 \hbar \int_{\frac{\rho^2}{2}}^\infty\! dt\, e^{-\frac t \hbar} (\tau -1)^{-\lambda-\frac 12} \hat \phi_-(\tau-1) \nn \\
 & = \rho^{\lambda } e^{-\frac{\rho^2}{4\hbar}} \Big(e^{i \pi (\lambda+\frac 12)} - e^{-i \pi (\lambda+\frac 12)} \Big) \kappa(\lambda)
\frac{2^{-\lambda-\frac 12}}{\hbar} \int_{0}^\infty\! dt\, e^{-\frac t \hbar} t^{-\lambda-\frac 12} \hat \phi_-(\tau) \nn \\
& = \frac{i  \sqrt{2\pi} \hbar^{-\lambda-\frac 12}}{\Gamma(1+\lambda)} \psi_-\,, \qquad z\in \mathbb{R}_+\,,
\label{ODQG7}
\end{align}
where in the second identity the change of variable of integration $t\rightarrow t+\rho^2/2$ has been performed.

When $z$ is real negative \eqref{eq:ODQG6b} still applies, but now $z=\rho e^{\pm i \pi}$. The choice of phase depends on how we split the Stokes line and the branch-cut
for $z<0$. Taking $z=\rho e^{i \pi}$ means that if we rotate counterclockwise from the positive real axis we first encounter the Stokes line and then the branch-cut.
Viceversa for  $z=\rho e^{-i \pi}$. For definiteness we choose the first assignment. With respect to the previous analysis, an extra phase is acquired in rewriting
\be
(\rho e^{i \pi})^{-\lambda-1} \rho^{2\lambda+1} = (\rho e^{i \pi})^{\lambda}e^{-2i \pi (\lambda+\frac 12)}\,,
\label{ODQG7b}
\ee
so that 
\be
(s_{0^+} -s_{0^-})\widetilde \psi_+  = \frac{i  \sqrt{2\pi} e^{-2i \pi (\lambda+\frac 12)} \hbar^{-\lambda-\frac 12}}{\Gamma(1+\lambda)} \psi_-\,, \qquad z\in \mathbb{R}_-\,.
\label{ODQG7a}
\ee
Consider now $(s_{\pi^+} - s_{\pi^-})\widetilde\psi_-$. 
We determine again $a_\pm$ by demanding that the non-analytic behaviour of $\hat \phi_-^{a_-}$ around $\tau=-1$ is given by $\hat \phi_+^{a_+}$ around $\tau=0$. 
We find $a_- = 0$, $a_+ = \lambda + 1/2$, resulting in
\be
\hat \phi_-(\tau)\Big|_{\tau=-1} =\widetilde \kappa(\lambda) (1+\tau)^{\lambda+\frac 1 2}\hat \phi_+(\tau+1) + \text{regular}\,,
\label{eq:ODQG12}
\ee
where $\hat \phi_- \equiv \hat \phi_-^{0}$,  $\hat \phi_+ \equiv \hat \phi_+^{\lambda+1/2}$, and
\be
\widetilde \kappa(\lambda) =  \frac{\Gamma\Big(-\lambda-\frac 12\Big)\Gamma\Big(\lambda+\frac 32\Big)}{\Gamma\Big(\frac{1-\lambda}2\Big) \Gamma\Big(-\frac \lambda2\Big)}\,.
\ee
For $z$ pure imaginary positive, $z = \rho \, \exp(i \frac{\pi}2)$,
we get
\begin{align}
(s_{\pi^+} -s_{\pi^-}) \widetilde\psi_- & = z^{\lambda} e^{-\frac{z^2}{4\hbar}} \Big(e^{-i \pi (\lambda+\frac 12)} - e^{i \pi (\lambda+\frac 12)} \Big)\widetilde\kappa(\lambda)
\frac 1 \hbar \int_{-\frac{z^2}{2}}^\infty\! dt\, e^{-\frac t \hbar} (\tau +1)^{\lambda+\frac 12}  \hat \phi_+(\tau+1) \nn \\
 & =  ( \rho e^{i \frac{\pi}2})^{\lambda}  \rho^{- \lambda -1} e^{+\frac{z^2}{4\hbar}} \Big(e^{-i \pi (\lambda+\frac 12)} - e^{i \pi (\lambda+\frac 12)} \Big)\widetilde \kappa(\lambda)
\frac 1 \hbar \int_{0}^\infty\! dt\, e^{-\frac t \hbar} t^{\lambda+\frac 12}  \hat \phi_+(\tau) \nn  \\
& =  \frac{-e^{i \pi \lambda} \sqrt{2\pi} \hbar^{\lambda+\frac 12}}{\Gamma(-\lambda)} \psi_+\,, \qquad {\rm Im}\; z > 0\,.
\label{ODQG8}
\end{align}
When $z$ is pure imaginary negative, $z = \rho \, \exp(-i \frac{\pi}2)$, we get
\be
(s_{\pi^+} -s_{\pi^-}) \widetilde\psi_-  =  \frac{e^{-i \pi \lambda}  \sqrt{2\pi}\hbar^{\lambda+\frac 12}}{\Gamma(-\lambda)} \psi_+\,, \qquad {\rm Im}\; z <  0\,.
\label{ODQG9}
\ee
In total we have 5 different connection matrices, as depicted in fig.\ref{fig:DQTP}:
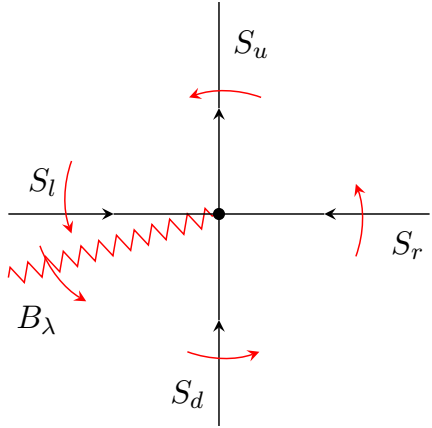
\begin{figure}[t!]
    \centering
    \raisebox{-3.em}
 {\scalebox{1.4}{
    \begin{tikzpicture}

\draw[snake=zigzag,segment length=5pt, red] (1,2.4)-- (3,3); 

   \filldraw[fill=black, draw=black]  (3,3) circle (0.05 cm);
      \draw [stealth-] (4,3) - - (5,3);
            \draw (3,3) - - (4,3);
          \node[right] at (4.5,2.7) {\scalebox{.8}{$S_r $}}; 
              \draw [-stealth,red] (4.3,2.6) arc (-20:20:10mm);
           
                      \draw [-stealth] (1,3) - - (2,3);
            \draw (2,3) - - (3,3);
                    \node[left] at (1.6,3.3) {\scalebox{.8}{$S_l $}}; 
          \draw [-stealth,red] (1.6,3.5) arc (160:200:10mm);
                            \draw [-stealth] (3,3) - - (3,4);
            \draw (3,4) - - (3,5);
                  \node[right] at (3.,4.6) {\scalebox{.8}{$S_u $}}; 
           \draw [-stealth,red] (3.4,4.1) arc (70:110:10mm);

                          \draw  (3,3) - - (3,2);
            \draw [stealth-] (3,2) - - (3,1);
                  \node[left] at (3.,1.3) {\scalebox{.8}{$S_d $}}; 
                          \draw [-stealth,red] (2.7,1.7) arc (250:290:10mm);
  
                  \node[left] at (1.6,2.) {\scalebox{.8}{$B_\lambda$}}; 
         \draw [-stealth,red] (1.3,2.7) arc (200:240:10mm);

 \end{tikzpicture}}}
    \caption{Connection matrices (including the branch-cut) for the Weber equation \eqref{ODQG1} (deformed quadratic turning point).}
	\label{fig:DQTP}

\end{figure} 

\begin{align}
S_r  & = \left(\begin{array}{cc}
1 &  0  \\
 \frac{i  \sqrt{2\pi}\hbar^{-\lambda-\frac 12}}{\Gamma(1+\lambda)} & 1
\end{array}\right) \,, \qquad \qquad \quad
S_u = \left(\begin{array}{cc}
1 &  \frac{-e^{i \pi \lambda}  \sqrt{2\pi}\hbar^{\lambda+\frac 12}}{\Gamma(-\lambda)}    \label{eq:ODQG9a} \\
0 & 1
\end{array}\right) \,, \\
S_l  & = \left(\begin{array}{cc}
1 & 0  \\
 -\frac{i  \sqrt{2\pi} e^{-2i \pi \lambda}\hbar^{-\lambda-\frac 12}}{\Gamma(1+\lambda)} & 1
\end{array}\right) \,, \qquad 
S_d = \left(\begin{array}{cc}
1 &  \frac{e^{-i \pi \lambda}  \sqrt{2\pi}\hbar^{\lambda+\frac 12}}{\Gamma(-\lambda)}  \\
0  & 1
\end{array}\right) \,, \qquad 
B_\lambda  =\left(\begin{array}{cc}
e^{-2i \pi \lambda} & 0  \\
0 & e^{2i \pi \lambda} 
\end{array}\right) \,. \nn
\end{align}
The subscripts $r$, $u$, $l$ and $d$ refer to the four Stokes lines in the positive real axis ($r$), positive imaginary axis $(u)$, negative real axis ($l$) and negative imaginary axis $(d)$ respectively. 
For simplicity we omit to write their dependence on $\lambda$.
$B_\lambda$ is the monodromy matrix due to the branch-cut at $z=0$. 
Solutions of the differential equations are analytic over the complex plane and so under a full rotation over the complex $z$-plane $\psi_\pm \rightarrow \psi_\pm$.
Indeed we have
\be
S_r S_d B_\lambda S_l S_u = I\,.
\label{eq:ODQG15}
\ee
As a sanity check we reproduce the pure case by setting $\lambda=-1/2$, in which case we have 
\be
S_r  =  S_l = S_-^{(2)} \,, \qquad  S_u = S_d = S_+^{(2)}\,, \qquad B_\lambda= - I\,,
\ee
where $S_\pm^{(2)}$  are the connection matrices \eqref{ODGp14a} for $q=2$. 
Equation \eqref{eq:ODQG15} and the $\lambda=-1/2$ match provide a quite non-trivial consistency check of the connection matrices \eqref{eq:ODQG9a}.
Note that \eqref{eq:ODQG15} depends on the location of the branch-cut. For instance, by changing
the order of the Stokes line and branch-cut for negative $z$, \eqref{ODQG7b} changes in 
\be
(\rho e^{-i \pi})^{-\lambda-1} \rho^{2\lambda+1} = (\rho e^{-i \pi})^{\lambda}e^{2i \pi (\lambda+\frac 12)}\,,
\label{ODQG715a}
\ee
and correspondingly
\be
S_l \rightarrow  S_l'   = \left(\begin{array}{cc}
1 & 0   \\
 -\frac{i \sqrt{2\pi} e^{2i \pi \lambda}\hbar^{-\lambda-\frac 12}}{\Gamma(1+\lambda)} & 1
\end{array}\right) = B_\lambda  S_l B_\lambda^{-1} \,,
\ee
so that 
\be
 S_u  B_\lambda   S_l'  S_d  S_r = I\,.
\label{eq:ODQG15b}
\ee
We recognize that the factors $\sqrt{2\pi} \hbar^{\lambda+1/2}/\Gamma(-\lambda)$ and $\sqrt{2\pi} \hbar^{-\lambda-1/2}/\Gamma(\lambda+1)$ appearing respectively in $S_{u,d}$
and $S_{r,l}$ precisely reproduce the similar factors in \eqref{eq:adcAsASG} and \eqref{stokes-pi}, upon identifying $\lambda$ with $t$.
In other words, when two simple turning points $A$ and $B$ collapse, the non-perturbative combination of Voros symbols $\sqrt{a_{DC}}$ and $\sqrt{a_{DC}/a_{AB}}(1+a_{AB})$
acquire a universal model-independent contribution which can be reinterpreted as a factor entering the connection matrix for the Weber equation \eqref{ODQG1}.
The remaining model-dependent terms, parametrized e.g. in the formal power series \eqref{eq:adcAsASG} by the constants $c_{1,2}$ and by the coefficients of the asymptotic series in $\hbar$ hidden in the ${\cal O}(\hbar)$ terms, can be fixed by e.g. using the so called uniform WKB method. This method was introduced back in \cite{Langer,Cherry,PhysRev.91.174} and more recently reconsidered, see e.g. \cite{Alvarez_2000} for anharmonic potentials and \cite{marino_2021} for a pedagogical introduction. In our context it essentially consists in finding a change of coordinates
which brings the original Schr\"odinger equation \eqref{eq:SchroC} to the Weber form \eqref{ODQG1}.  See also \cite{AKT2,Takei1995OnTC} for a closely related way to fix such terms.
We will not discuss here how to determine the model-dependent factors using such techniques, since we can compute them by working with simple turning points in the limit $e_0\rightarrow 0$, as explained in the main text. 

\section{Transseries in the pure quartic anharmonic}
\label{app:purequartic}

In this appendix we discuss in some detail the quartic anharmonic potential with no mass term as an example of a model where perturbation theory in $\hbar$ is ill-defined.
Notably this is the first model where EWKB has been applied  \cite{bpv1,voros-quartic}.
A simple scaling argument shows that the energy levels have the form $E_n = \hbar^{4/3} \gamma_n$, where $\gamma_n$ are real numbers independent of $\hbar$. 
Although the $\hbar$-expansion does not exist, we can consider a semiclassical expansion in $1/n$. 
We show below that energy eigenvalues $E_n$ are given by a transseries in $1/n$ and $\exp(-\pi n)$, and we study its convergence properties.
Since the quartic model can be easily obtained from EPT, as in \eqref{eq:ept-eg}, this also illustrates how an observable can be packaged in a dramatically different way by EPT.

Like in the massive quartic anharmonic model discussed in the main text, the pure quartic admit eight different wedges in the $\hbar$ complex plane delimited by Stokes jumps.
Unlike the massive quartic, however, the Stokes jump separating region I from region II is independent of $E$ and occurs at 
$\arctan\left( \frac{ \Pi_{0,BA}(E)}{i \Pi_{0,CD}(E)}\right) =\frac{\pi}{4}$. No sweet spot can be defined and hence we 
focus our analysis just on the region I delimiting the real $\hbar$-axis. The EQC in this region can be read from the first line of \eqref{eqcs-reduced}
and can be written by simple manipulations of periods as 
\begin{equation}
		\label{eqPQ:quantization_condition}
		\sqrt{a_{BA}^{\text{I}}} = \pm i - \sqrt{a_{DC}^{\text{I}}} \,,
	\end{equation}
	for even/odd levels. The two periods can be written purely as functions of $x\equiv \hbar/E^{3/4}$ in terms of a single function $R$:
	\begin{equation}
		\sqrt{a_{BA}^{\text{I}}} = e^{iR(x)},\qquad \sqrt{a_{DC}^{\text{I}}} = e^{-iR(ix)}\,.
\end{equation}
The nontrivial relation between the perturbative and nonperturbative cycle, which allows us to express both in terms of the same function $R(x)$, can be seen at each order in $n$ noting that the nonperturbative cycle is obtained from a complex rotation $z\rightarrow iz$ of the perturbative one, while the integrand has the form $f_k(z^4)$ for $k$ even, $z^2f_k(z^4)$ for $k$ odd, thus picking up a minus sign for odd terms only.

Since the Voros symbol $a_{BA}$ is not Borel resummable, so will be the formal asymptotic series
\begin{equation}
    \widetilde R= \sum_{k=0} a_k x^{2k-1}
    \label{app:definition_Rtilde}
\end{equation}
associated to $R$ (see \cite{gmz} for an explicit analysis).
Neglecting the nonperturbative contribution given by $a_{DC}^{\text{I}}$, the EQC \eqref{eqPQ:quantization_condition} implies
\begin{equation}
\label{app:quantization_condition_perturbative}
		\widetilde R(x) = \nu\,, \qquad  \nu\equiv \pi\left(n+\frac{1}{2}\right)\,.
	\end{equation}
Inverting the above formal series, we get
\be
\widetilde x_{\text{P}}(\nu) = \widetilde R^{-1}(\nu)\sim \frac{a_0}{\nu}+\frac{a_0a_1}{\nu^3}+\dots \,.
\label{eq:asymptxtilde}
\ee
As we will see below, the presence of the nonperturbative cycle turns the asymptotic series $\widetilde x_{\text{P}}$ in \eqref{eq:asymptxtilde} in a transseries $\widetilde x(\nu)$ defined as
\begin{equation}
	\widetilde x= \sum_{j,k=0}^\infty c_{j,k} \frac{\epsilon^j}{\nu^{2k+1}}\, , \qquad \epsilon\equiv e^{-(1+i)\nu} \,,
 \label{app:xTS}
\end{equation}
where $c_{0,0}=a_0$, $c_{0,1}=a_0 a_1$, $\ldots$, from which the resummed energy eigenvalues can be obtained through Borel resummation as
\be
E_n = \hbar^{\frac{4}{3}} s\left(\widetilde x\right)^{-\frac{4}{3}}= \hbar^{\frac{4}{3}} \left(\sum_{j=0}^\infty(\mp i)^{j} e^{-j\pi(n+\frac{1}{2})}s_+\left(\sum_{k=0}^\infty \frac{c_{j,k}}{\left(\pi(n+\frac{1}{2})\right)^{2k+1}}\right)\right)^{-\frac{4}{3}}\,,
\label{app:EnDefTS}
\ee
where $s_+$ indicates a lateral Borel resummation slightly above the positive $\hbar$ real axis (region I) and $\mp$ refers respectively to even and odd $n$.
The appearance of imaginary terms in the physical energy eigenvalues should not surprise. They are there precisely to compensate imaginary terms arising from the lateral Borel resummation in the spirit of resurgence. Naturally, we require that the sum over $j$ results in a finite number. It has been observed in the literature on resurgence, see e.g. \cite{ecalle, costin-conv}, that in many cases the sum over $j$ is convergent for sufficiently large $n$, or $\nu$. 
A related observation is that if we formally reorganise $\widetilde{x}$ as
\begin{equation}
\widetilde x =  \sum\limits_{k=0}\frac{\widetilde{S}_k}{\nu^{2k+1}},\qquad \widetilde{S}_k\equiv\sum_{j=0}^\infty c_{j,k}\, e^{-j(1+i) \nu},
 \label{app:fullTS}
\end{equation}
the partial sums $\widetilde S_k$ can themselves be convergent. 
For example, in \cite{costin-rad} it is shown that a broad class of non-linear ODEs admit formal solutions in terms of transseries where the analogous partial series $\widetilde{S}_k$ are all convergent within the radius of convergence of $\widetilde{S}_0$. 
We show below that this also happens with the partial series $\widetilde{S}_k$ in \eqref{app:fullTS}. 
We conjecture that this is evidence of the fact that the transseries \eqref{app:xTS} (i.e. the sum over $j$ in \eqref{app:EnDefTS}) is convergent in approximately the same region.

We now show that \eqref{app:xTS} applies and study the convergent properties of the factors $\widetilde{S}_k$ in \eqref{app:fullTS}.
We massage \eqref{eqPQ:quantization_condition} into
	\begin{equation}
		\label{eqPQ:quantization_condition_massaged}
		e^{iR(x)} = e^{i\nu}\left(1-e^{-iR(ix)-i\nu}\right)\Rightarrow R(x) = \nu -i\ln\left(1-e^{\nu-iR(ix)}e^{-(1+i)\nu}\right) \,,
\end{equation}
where $e^{\nu-iR(ix)}\sim\OO(1)$ for parametrically small $x$ (or large $\nu$). 
If we assume that $\widetilde{x}$ is known up to order $\epsilon^N$ and to sufficiently high order in $\nu^{-1}$, then $R(ix)$ and its exponential are known to $\OO(\epsilon^N)$. The additional factor $\epsilon$ inside the logarithm then allows us to compute to order $\epsilon^{N+1}$. Iterating the process, \eqref{eqPQ:quantization_condition_massaged} is satisfied to each order by the transseries ansatz \eqref{app:xTS}.
		
Let us denote the radius of convergence of $\widetilde{S}_k(\epsilon)$ as $r_k$. To obtain it, first notice that the coefficient of $1/\nu^{2k+1}$ is determined from the $\widetilde{R}$ series up to the term $a_k$. Thus to get the series $\widetilde S_0$ it is sufficient to truncate $\widetilde R(x)$ to $\frac{a_0}{x}$, neglecting $\OO(x)$. We leave higher order terms for later.
	We replace $-i\widetilde R(ix),-\widetilde{R}(x) \rightarrow -\mathcal{R}$ and \eqref{eqPQ:quantization_condition_massaged} gives
 \be
w=1-\epsilon e^{i\ln w},\qquad \qquad w\equiv e^{i(\mathcal{R}-\nu)} \,.
\label{eqPQ:eqW}
\ee
Note that the $x$ dependence is now completely implicit. We can see from \eqref{eqPQ:eqW} that all derivatives of $w(\epsilon)$ are bounded around the origin and hence analyticity is guaranteed at $\epsilon=0$. Due to the logarithm, any $\bar \epsilon$ such that $w(\bar{\epsilon})=0$, must be at the edge of the radius of convergence of the series $w(\epsilon)$ defined by \eqref{eqPQ:eqW}. Noting that $w(0)=1$, we now consider all paths $\gamma\in\Gamma$, in the complex $w$ plane, going from $w=1$ to $w=0$ and the corresponding value of $|\epsilon|$ along each path. The radius of convergence will be $\min\limits_{\gamma\in\Gamma}\max\limits_{w\in\gamma}|\epsilon(w)|$, because if it were greater then we could construct the curves $\epsilon(t)$, $w(\epsilon(t))$ such that $w$ approaches $0$ and $\epsilon(t)$ is always in the radius of convergence. Conversely, there are no other singularities in \eqref{eqPQ:eqW} obstructing convergence.

\begin{figure}[t!]
\centering
\includegraphics[width=0.45\linewidth]{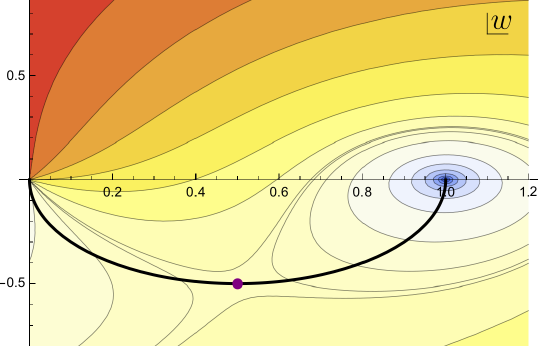}
\caption{Level curves of the function $|\epsilon(w)|^2$. In black an optimal curve 
$\gamma^*$. The purple dot is where the maximum value of $|\epsilon|$ is achieved.}
\label{figure:app_epsilon}
\end{figure}
 
Letting $w(t)=L(t)e^{i\theta(t)}$, \eqref{eqPQ:eqW} gives
\begin{equation}
|\epsilon|^2 = e^{2\theta}\left(1+L^2-2L\cos\theta\right) \,.
\end{equation}
Any curve that connects $w=0$ to $w=1$ must pass by the line~$\textrm{Re}(w)=1/2$, along which $|\epsilon|^2$ has a single global minimum, see figure \ref{figure:app_epsilon}. Since $\nabla|\epsilon|^2$ vanishes at $L=\frac{1}{\sqrt{2}}$ and $\theta=-\frac{\pi}{4}$, the minimum value of $|\epsilon|^2$ in this line is at $w=\frac{1-i}{2}$, and thus $\min\limits_{\gamma\in\Gamma}\max\limits_{w\in\gamma}|\epsilon(w)|\geq|\epsilon\left(\frac{1-i}{2}\right)|$. To reduce the bound to an equality, consider the curve $\gamma^*$ such that $L=\cos\theta$; $\max\limits_{w\in\gamma^*}|\epsilon|$ is easily computed to be exactly at~$w=\frac{1-i}{2}$. Thus, the radius of convergence $r_0$ of the series $\widetilde S_0(\epsilon)$ is 
\be
r_0\equiv \max\limits_{w\in\gamma^*}|\epsilon| = \frac{e^{-\pi/4}}{\sqrt{2}}\,,
\ee
a result we numerically confirmed. Importantly, all quantized values of $\nu$ lie inside this radius.
	
We now want to extend the result to $\widetilde S_{k>0}$. Recalling that to determine $\widetilde S_k$ we only need finitely many terms in $\nu^{-1}$, let us truncate $\widetilde x$ as it is defined in equation \eqref{app:fullTS},
\begin{equation}
    \widetilde x^{(k+1)}\equiv\sum\limits_{k'=0}^{k+1}\frac{\widetilde{S}_{k'}(\epsilon)}{\nu^{2k'+1}} = \widetilde x^{(k)}+\frac{\widetilde{S}_{k+1}}{\nu^{2k+3}}.
\end{equation}
Plugging this ansatz for $\widetilde x$ into $\widetilde R(x)$ (equation \eqref{app:definition_Rtilde}) and then $\widetilde R(x)$ into equation~\eqref{eqPQ:quantization_condition_massaged}, we obtain $\widetilde{S}_{k+1}$. It is expressed as a sum of products of the previously encountered $\{\widetilde{S}_i(\epsilon)\}_{i\le k}$ so, 
by Merten's theorem,
the radius of convergence $r_{k+1}$ of $\widetilde S_{k+1}$ is $r_{k+1}\ge\min\limits_{i\le k} \{r_i\}$.

\bibliographystyle{JHEP}
\bibliography{Refs}

\providecommand{\href}[2]{#2}\begingroup\raggedright\begin{thebibliography}{10}

\bibitem{Bender:1969si}
C.~M. Bender and T.~T. Wu, {\it {Anharmonic oscillator}},  {\em Phys. Rev.}
  {\bf 184} (1969) 1231--1260.

\bibitem{Bender:1973rz}
C.~M. Bender and T.~T. Wu, {\it {Anharmonic oscillator. 2: A Study of
  perturbation theory in large order}},  {\em Phys. Rev. D} {\bf 7} (1973)
  1620--1636.

\bibitem{Wentzel:1926aor}
G.~Wentzel, {\it {Eine Verallgemeinerung der Quantenbedingungen f\"ur die
  Zwecke der Wellenmechanik}},  {\em Z. Phys.} {\bf 38} (1926), no.~6 518--529.

\bibitem{Kramers:1926njj}
H.~A. Kramers, {\it {Wellenmechanik und halbzahlige Quantisierung}},  {\em Z.
  Phys.} {\bf 39} (1926), no.~10 828--840.

\bibitem{Brillouin:1926blg}
L.~Brillouin, {\it {La m\'ecanique ondulatoire de Schr\"odinger; une m\'ethode
  g\'en\'erale de resolution par approximations successives}},  {\em Compt.
  Rend. Hebd. Seances Acad. Sci.} {\bf 183} (1926), no.~1 24--26.

\bibitem{voros-quartic}
A.~Voros, {\it The return of the quartic oscillator. {T}he complex {WKB}
  method},  {\em Annales de l'I.H.P. Physique Th\'eorique} {\bf 39} (1983),
  no.~3 211--338.

\bibitem{silverstone}
H.~J. Silverstone, {\it {JWKB} connection-formula problem revisited via {B}orel
  summation},  {\em Phys. Rev. Lett.} {\bf 55} (1985), no.~23 2523.

\bibitem{ecalle}
J.~{\'E}calle, {\it Les fonctions r{\'e}surgentes},  {\em Publ. math.
  d'Orsay/Univ. de Paris, Dep. de math.} (1981).

\bibitem{Balian:1974ah}
R.~Balian and C.~Bloch, {\it {Solution of the Schrodinger Equation in Terms of
  Classical Paths}},  {\em Annals Phys.} {\bf 85} (1974) 514.

\bibitem{bpv1}
R.~Balian, G.~Parisi, and A.~Voros, {\it Discrepancies from asymptotic series
  and their relation to complex classical trajectories},  {\em Phys. Rev.
  Lett.} {\bf 41} (Oct, 1978) 1141--1144.

\bibitem{AKT1}
T.~Aoki, T.~Kawai, and Y.~Takei, {\it The bender-wu analysis and the voros
  theory},  in {\em ICM-90 Satellite Conference Proceedings} (M.~Kashiwara and
  T.~Miwa, eds.), (Tokyo), pp.~1--29, Springer Japan, 1991.

\bibitem{reshyper}
H.~Dillinger, E.~Delabaere, and F.~d.~r. Pham, {\it R\'esurgence de {V}oros et
  p\'eriodes des courbes hyperelliptiques},  in {\em Annales de l'institut
  Fourier}, vol.~43, p.~163, 1993.

\bibitem{ddpham}
E.~Delabaere, H.~Dillinger, and F.~Pham, {\it Exact semiclassical expansions
  for one-dimensional quantum oscillators},  {\em J. Math. Phys.} {\bf 38}
  (1997), no.~12 6126--6184.

\bibitem{dpham}
E.~Delabaere and F.~Pham, {\it Resurgent methods in semi-classical
  asymptotics},  {\em Annales de l'IHP} {\bf 71} (1999) 1--94.

\bibitem{AKT2}
T.~Aoki, T.~Kawai, and Y.~Takei, {\it The bender wu analysis and the voros
  theory ii},  {\em Advanced Studies in Pure Mathematics} {\bf 54} (2009)
  19--94.

\bibitem{Zinn-Justin:1982hva}
J.~Zinn-Justin, {\it {Multi - Instanton Contributions in Quantum Mechanics.
  2.}},  {\em Nucl. Phys. B} {\bf 218} (1983) 333--348.

\bibitem{Gaiotto:2009hg}
D.~Gaiotto, G.~W. Moore, and A.~Neitzke, {\it {Wall-crossing, Hitchin systems,
  and the WKB approximation}},  {\em Adv. Math.} {\bf 234} (2013) 239--403,
  [\href{http://arxiv.org/abs/0907.3987}{{\tt arXiv:0907.3987}}].

\bibitem{Mironov:2009uv}
A.~Mironov and A.~Morozov, {\it {Nekrasov Functions and Exact Bohr-Zommerfeld
  Integrals}},  {\em JHEP} {\bf 04} (2010) 040,
  [\href{http://arxiv.org/abs/0910.5670}{{\tt arXiv:0910.5670}}].

\bibitem{cm-ha}
S.~Codesido and M.~Mari\~no, {\it {Holomorphic Anomaly and Quantum Mechanics}},
   {\em J. Phys.} {\bf A51} (2018), no.~5 055402,
  [\href{http://arxiv.org/abs/1612.07687}{{\tt arXiv:1612.07687}}].

\bibitem{coms}
S.~Codesido, M.~Mari\~no, and R.~Schiappa, {\it {Non-Perturbative Quantum
  Mechanics from Non-Perturbative Strings}},
  \href{http://arxiv.org/abs/1712.02603}{{\tt arXiv:1712.02603}}.

\bibitem{Ito:2018eon}
K.~Ito, M.~Mari\~no, and H.~Shu, {\it {TBA equations and resurgent Quantum
  Mechanics}},  {\em JHEP} {\bf 01} (2019) 228,
  [\href{http://arxiv.org/abs/1811.04812}{{\tt arXiv:1811.04812}}].

\bibitem{dt}
P.~Dorey and R.~Tateo, {\it {Anharmonic oscillators, the thermodynamic Bethe
  ansatz, and nonlinear integral equations}},  {\em J.Phys.} {\bf A32} (1999)
  L419--L425, [\href{http://arxiv.org/abs/hep-th/9812211}{{\tt
  hep-th/9812211}}].

\bibitem{Gu:2022fss}
J.~Gu and M.~Marino, {\it {On the resurgent structure of quantum periods}},
  {\em SciPost Phys.} {\bf 15} (2023) 035,
  [\href{http://arxiv.org/abs/2211.03871}{{\tt arXiv:2211.03871}}].

\bibitem{dunne-unsal}
G.~V. Dunne and M.~\"Unsal, {\it {Uniform WKB, multi-instantons, and resurgent
  trans-series}},  {\em Phys. Rev.} {\bf D89} (2014), no.~10 105009,
  [\href{http://arxiv.org/abs/1401.5202}{{\tt arXiv:1401.5202}}].

\bibitem{Basar:2017hpr}
G.~Basar, G.~V. Dunne, and M.~Unsal, {\it {Quantum Geometry of Resurgent
  Perturbative/Nonperturbative Relations}},  {\em JHEP} {\bf 05} (2017) 087,
  [\href{http://arxiv.org/abs/1701.06572}{{\tt arXiv:1701.06572}}].

\bibitem{Sueishi:2020rug}
N.~Sueishi, S.~Kamata, T.~Misumi, and M.~\"Unsal, {\it {On exact-WKB analysis,
  resurgent structure, and quantization conditions}},  {\em JHEP} {\bf 12}
  (2020) 114, [\href{http://arxiv.org/abs/2008.00379}{{\tt arXiv:2008.00379}}].

\bibitem{Sueishi:2021xti}
N.~Sueishi, S.~Kamata, T.~Misumi, and M.~\"Unsal, {\it {Exact-WKB, complete
  resurgent structure, and mixed anomaly in quantum mechanics on S$^{1}$}},
  {\em JHEP} {\bf 07} (2021) 096, [\href{http://arxiv.org/abs/2103.06586}{{\tt
  arXiv:2103.06586}}].

\bibitem{Kamata:2021jrs}
S.~Kamata, T.~Misumi, N.~Sueishi, and M.~\"Unsal, {\it {Exact WKB analysis for
  SUSY and quantum deformed potentials: Quantum mechanics with Grassmann fields
  and Wess-Zumino terms}},  {\em Phys. Rev. D} {\bf 107} (2023), no.~4 045019,
  [\href{http://arxiv.org/abs/2111.05922}{{\tt arXiv:2111.05922}}].

\bibitem{Serone:2016qog}
M.~Serone, G.~Spada, and G.~Villadoro, {\it {Instantons from Perturbation
  Theory}},  {\em Phys. Rev. D} {\bf 96} (2017), no.~2 021701,
  [\href{http://arxiv.org/abs/1612.04376}{{\tt arXiv:1612.04376}}].

\bibitem{power}
M.~Serone, G.~Spada, and G.~Villadoro, {\it {The Power of Perturbation
  Theory}},  {\em JHEP} {\bf 05} (2017) 056,
  [\href{http://arxiv.org/abs/1702.04148}{{\tt arXiv:1702.04148}}].

\bibitem{INcluster}
K.~Iwaki and T.~Nakanishi, {\it Exact wkb analysis and cluster algebras},  {\em
  Journal of Physics A: Mathematical and Theoretical} {\bf 47} (nov, 2014)
  474009, [\href{http://arxiv.org/abs/1401.7094}{{\tt arXiv:1401.7094}}].

\bibitem{marino_2021}
M.~Mari\~no, {\em {Advanced Topics in Quantum Mechanics}}.
\newblock Cambridge University Press, 2021.

\bibitem{DELABAERE1997180}
E.~Delabaere and F.~Pham, {\it Unfolding the quartic oscillator},  {\em Annals
  of Physics} {\bf 261} (1997), no.~2 180--218.

\bibitem{Graffi:1970erh}
S.~Graffi, V.~Grecchi, and B.~Simon, {\it {Borel summability: Application to
  the anharmonic oscillator}},  {\em Phys. Lett. B} {\bf 32} (1970) 631--634.

\bibitem{Dorigoni:2014hea}
D.~Dorigoni, {\it {An Introduction to Resurgence, Trans-Series and Alien
  Calculus}},  {\em Annals Phys.} {\bf 409} (2019) 167914,
  [\href{http://arxiv.org/abs/1411.3585}{{\tt arXiv:1411.3585}}].

\bibitem{abs}
I.~Aniceto, G.~Basar, and R.~Schiappa, {\it {A Primer on Resurgent Transseries
  and Their Asymptotics}},  {\em Phys. Rept.} {\bf 809} (2019) 1--135,
  [\href{http://arxiv.org/abs/1802.10441}{{\tt arXiv:1802.10441}}].

\bibitem{Sauzin}
D.~Sauzin, {\it {Resurgent functions and splitting problems}},  {\em Res. Inst.
  Math. Sci.} {\bf 1493} (2006) 48, [\href{http://arxiv.org/abs/0706.0137}{{\tt
  arXiv:0706.0137}}].

\bibitem{huang-dif-op}
M.-x. Huang, {\it {On Gauge Theory and Topological String in
  Nekrasov-Shatashvili Limit}},  {\em JHEP} {\bf 06} (2012) 152,
  [\href{http://arxiv.org/abs/1205.3652}{{\tt arXiv:1205.3652}}].

\bibitem{Emery:2020qqu}
Y.~Emery, {\it {TBA equations and quantization conditions}},  {\em JHEP} {\bf
  07} (2021) 171, [\href{http://arxiv.org/abs/2008.13680}{{\tt
  arXiv:2008.13680}}].

\bibitem{Jaeckel:2018tdj}
J.~Jaeckel and S.~Schenk, {\it {Exploring high multiplicity amplitudes: The
  quantum mechanics analogue of the spontaneously broken case}},  {\em Phys.
  Rev. D} {\bf 99} (2019), no.~5 056010,
  [\href{http://arxiv.org/abs/1811.12116}{{\tt arXiv:1811.12116}}].

\bibitem{serone2}
M.~Serone, G.~Spada, and G.~Villadoro, {\it {$\lambda \phi_2^4$ theory Part II.
  the broken phase beyond NNNN(NNNN)LO}},  {\em JHEP} {\bf 05} (2019) 047,
  [\href{http://arxiv.org/abs/1901.05023}{{\tt arXiv:1901.05023}}].

\bibitem{Langer}
R.~E. Langer, {\it The asymptotic solutions of ordinary linear differential
  equations of the second order, with special reference to a turning point},
  {\em Transactions of the American Mathematical Society} {\bf 67} (1949),
  no.~2 461--490.

\bibitem{Cherry}
T.~M. Cherry, {\it Uniform asymptotic formulae for functions with transition
  points},  {\em Transactions of the American Mathematical Society} {\bf 68}
  (1950) 224--257.

\bibitem{PhysRev.91.174}
S.~C. Miller and R.~H. Good, {\it A wkb-type approximation to the schr\"odinger
  equation},  {\em Phys. Rev.} {\bf 91} (Jul, 1953) 174--179.

\bibitem{Alvarez_2000}
G.~\'Alvarez and C.~Casares, {\it Uniform asymptotic and jwkb expansions for
  anharmonic oscillators},  {\em Journal of Physics A: Mathematical and
  General} {\bf 33} (apr, 2000) 2499.

\bibitem{Takei1995OnTC}
Y.~Takei, {\it On the connection formula for the first painleve equation : from
  the viewpoint of the exact wkb analysis(painleve transcendents and asymptotic
  analysis)},  1995.

\bibitem{gmz}
A.~Grassi, M.~Mari\~no, and S.~Zakany, {\it {Resumming the string perturbation
  series}},  {\em JHEP} {\bf 1505} (2015) 038,
  [\href{http://arxiv.org/abs/1405.4214}{{\tt arXiv:1405.4214}}].

\bibitem{costin-conv}
O.~Costin, {\it {Exponential asymptotics, transseries, and generalized Borel
  summation for analytic, nonlinear, rank-one systems of ordinary differential
  equations}},  {\em International Mathematics Research Notices} {\bf 1995}
  (04, 1995) 377--417,
  [\href{http://arxiv.org/abs/https://academic.oup.com/imrn/article-pdf/1995/8/377/6768513/1995-8-377.pdf}{{\tt
  https://academic.oup.com/imrn/article-pdf/1995/8/377/6768513/1995-8-377.pdf}}].

\bibitem{costin-rad}
O.~Costin and R.~D. Costin, {\it On the formation of singularities of solutions
  of nonlinear differential systems in antistokes directions},  {\em
  Inventiones mathematicae} {\bf 145} (2001), no.~3 425--485.

\end{thebibliography}\endgroup

\end{document}